\pdfoutput=1
\RequirePackage{fix-cm}

\documentclass[smallextended]{svjour3}       
\smartqed  
\usepackage[numbers]{natbib}

\usepackage{dblfloatfix}
\usepackage{arydshln}
\usepackage{xcolor}

\usepackage{balance}
\usepackage{alltt}
\usepackage{subcaption}

\usepackage{pifont}
\usepackage{graphicx}
\usepackage{colortbl}
\PassOptionsToPackage{dvipsnames}{xcolor}
\usepackage{rotating}
\usepackage{makecell}
\usepackage{tabularx}
\usepackage{booktabs}
\usepackage{wrapfig}
\usepackage{tikz}
\usetikzlibrary{angles}
\usepackage{multirow}
\usepackage{hyperref}

\usepackage{framed}
\usepackage[framemethod=tikz]{mdframed}
\usetikzlibrary{shadows}
\usepackage{enumitem}
\usepackage{graphics}
\usepackage[para,online,flushleft]{threeparttable}

\setlist[description]{leftmargin=1cm}

\usepackage{amsmath}

\usepackage{pgfplots}
\usetikzlibrary{patterns}
\usetikzlibrary{arrows}
\usetikzlibrary {positioning}
\usepackage{enumitem}
\setlist[itemize]{leftmargin=*}
\setlist[enumerate]{leftmargin=*}
\usepackage{makecell}
\usepackage{pifont}

\usepackage[linesnumbered,ruled,vlined]{algorithm2e}
\setlist{nolistsep} 

\SetKwProg{Fn}{Function}{}{}

\setlist[1]{itemsep=0pt}

\newmdenv[
tikzsetting= {fill=gray!10},
linewidth=1pt,
roundcorner=2pt,
shadow=false
]{myshadowbox}

\newcolumntype{P}[1]{>{\centering\arraybackslash}p{#1}}

\hypersetup{
    linkcolor=blue,
    filecolor=magenta,      
    urlcolor=cyan,
}

\newcommand{\bi}{\begin{itemize}[leftmargin=0.4cm]}
\newcommand{\ei}{\end{itemize}}
\newcommand{\be}{\begin{enumerate}[leftmargin=0.4cm]}
\newcommand{\ee}{\end{enumerate}}

\newcommand{\tbl}[1]{Table~\ref{tbl:#1}}

\usepackage{url}

\tikzstyle{thmbox} = [rectangle, rounded corners, draw=black, fill=gray!10]

\usepackage[tikz]{bclogo}
\usepackage{transparent}

\definecolor{cyan}{HTML}{2FF3E0}
\newenvironment{RQ}[1]%
{\noindent\begin{minipage}[c]{\linewidth}%
\begin{bclogo}[couleur=gray!30,%
                arrondi=0.1,%
                logo=\bctrombone,%
                ombre=true]{{\normalsize ~#1}}}%
{\end{bclogo}\end{minipage}\vspace{2mm}}

\newcommand*\colourcheck[1]{%
  \expandafter\newcommand\csname #1check\endcsname{\textcolor{#1}{\ding{51}}}%
}

\colourcheck{green}

\newcommand*\colourx[1]{%
  \expandafter\newcommand\csname #1x\endcsname{\textcolor{#1}{\ding{53}}}%
}
\colourx{red}

\usepackage{listings}
\usepackage{cellspace}
\begin{document}

\title{Less, but Stronger: On the Value of Strong Heuristics in Semi-supervised Learning \\ for Software Analytics}
\titlerunning{On the Value of Strong Heuristics in Semi-Supervised Learning}

\author{Huy Tu   \and
         Tim Menzies 
}

\institute{H. Tu,  and T. Menzies \at
              Department of Computer Science, \\
              North Carolina State University, \\
              Raleigh, USA \\
              \email{huyqtu7@gmail.com and timm@ieee.org}}





\date{Received: date / Accepted: date \vspace{-10pt}}

\maketitle

\begin{abstract}

In many domains, there are many examples and far fewer labels for those examples; e.g. we may have access to millions of lines of source code,
but access to only a handfuls of warnings about that code.
In those domains, semi-supervised learners (SSL) can extrapolate labels from
a small number of examples to the rest of the data.

Standard
SSL algorithms use ``weak'' knowledge (i.e. those not based on specific SE knowledge) such as (e.g.) co-train two learners and use good labels from one to train the other.

Another approach of SSL in software analytics is potentially use ``strong'' knowledge that use SE knowledge.
For example, an often-used heuristic in SE is that unusually large
artifacts contain undesired properties (e.g. more bugs).
This paper argues that  such   ``strong'' algorithms perform
better than those  standard, weaker, SSL algorithms.
We show this by learning   models
from   labels generated using weak SSL or our ``stronger'' FRUGAL
algorithm. 
In four domains
(distinguishing security-related  bug reports;
mitigating bias in decision making;
predicting issue close time; and 
(reducing false alarms in   static code warnings),
 FRUGAL required only 2.5\% of the data to be labeled
yet out-performed  standard semi-supervised
learners that relied on (e.g.) some domain independent graph theory
concepts.

Hence, for future work, we strongly recommend the use of strong
heuristics for semi-supervised learning for SE applications. To better support other researchers,  our scripts and data are on-line at
https://github.com/HuyTu7/FRUGAL.




\end{abstract}



\section{Introduction}\label{sec:introduction}

Standard software analytics  uses  supervised
learning  where models are learned from data
labelled with (e.g.) ``this is/is not a buggy commit''. Hence,
in standard analytics,
{\em data labeling} is a    vital initial task. 
But manual labelling can be very error prone~\cite{jitterbug, tu2020better, data_jinx, quality, herzig13_misclassifications}).
Standard practice   ~\cite{commitguru, Kim08changes,catolino17_jitmobile,nayrolles18_clever,mockus00changeskeys,kamei12_jit,hindle08_largecommits} is to label a commit  ``bug-fixing''
if the commit text has keywords (e.g.) ``bug'', ``fix'', ``error'', ``patch'', etc.
Such keyword-based approaches  often misclassify a commit (i.e., a false-positive like ``documentation updates and fixes'')~\cite{Vasilescu15github,Vasilescu18z,tu2020better}.
Worse still,   labelling can be very expensive; e.g. manually labelling  the  500   projects used
in one study~\cite{tu2020better} needs 39,000 hours (which, on a crowd-sourced
platform, would cost \$320,000).

Accordingly, this paper explores ways to better (and faster and cheaper and more
accurately) label examples.
We will constrast two approaches that use  {\em weak} or {\em strong} heuristics. In the AI literature~\cite{AI_4thedition}, ``weak heusitics'' are general purpose algorithms that make little-or-no use of domain knowledge. Examples of such weak heuristics algorithms are depth-first-search or simple theorem provers. 
Strong heuristics, on the other hand, exploit specific domain knowledge. For example, if you have lost your car keys, a weak depth-first search might lead t every room in the house. A better approach might be to first apply the strong heuristic ``look behind the couch cushions''. 

We think that this strong-versus-weak distinction is relevant and useful for software analytics. For example,
consider the problem of reducing the cost of data labelling:
\bi
\item
One solution to this approach is to use the weak heuristics of \S\ref{standardssl}; i.e.
cluster the data using some  algorithms,
label just a few items in each cluster, then use 
graph-theoretic methods to automatically propagate those labels
to nearby clusters.
\item 
Another approach, which we call FRUGAL,  makes no use of graph theory, nor co-training nor any of the other   techniques seen in standard AI semi-supervised  algorithms. Instead,
 our FRUGAL   tool is   dependent on the following strong SE   heuristic:
\ei
\begin{RQ}{
 Big things tend to have more problems.}
\end{RQ}  
That is, unusually large and general methods are more likely
to be buggy, take longer to resolve issues,
may hide more security bugs, may generate more
true positive static code warnings, etc.

This paper makes the case that strong heuristics are
valuable and generality across a broad range of SE tasks. To make that
case, we show here that we can have successfully apply FRUGAL's strong heuristic to:
\bi
\item
Distinguishing security-related  bug reports;
\item
Mitigating bias in decision making;
\item
Predicting issue close time; and 
\item
Reducing false alarms in   static code warnings.
\ei
 In all those studies,  our strong SE heuristic typically defeated
the weaker graph-theoretic heuristcs used by 
standard semi-supervised learning methods \citet{Shu2021HowTB,fair_smote,intrinsic_static}.
Better yet, these results were achieved   
after using merely   2.5\%  (i.e. $\frac{1}{40}$th) of the labels in the data.

We say that our results are interesting in two ways:
\bi
\item It shows that we can reduce the cost of labelling
data by a factor of $\frac{100}{2.5}=40$.
\item It shows that specific SE knowledge can be
more useful (for SE applications) than domain-independent
notions of co-training or label-spreading.
\ei
 \begin{table}[!t]
\vspace{-5pt}
\scriptsize
\renewcommand{\arraystretch}{1.1}
\caption{For self-admitted technical debt identification task, here are some examples of different labels from the original datasets curated by~\citet{maldonado2015detecting} and the updated datasets by~\citet{jitterbug}.} \vspace{-5pt}
\centering
\setlength\tabcolsep{4pt}
\begin{tabular}{p{.12\linewidth}|p{.52\linewidth}|P{.13\linewidth}|P{.13\linewidth}}
\textbf{Project} &  
\textbf{Comment Text} & 
\textbf{Original}  &
\textbf{Yu et al.'s} \\
 & & 
\textbf{Label~\cite{maldonado2015detecting}}  &
\textbf{Label~\cite{jitterbug}} \\
\hline
Apache Ant & 
\texttt{//TODO Test on other versions of weblogic //TODO add more attributes to the task, to take care of all jspc options //TODO Test on Unix} & 
no & 
yes \\
\hline
ArgoUML & 
\texttt{// skip backup files. This is actually a workaround for the cpp generator, which always creates backup files (it's a bug).}
 & no & yes \\
\hline
JFreeChart & 
\texttt{// FIXME: we've cloned the chart, but the dataset(s) aren't cloned and we should do that}  & 
no & 
yes \\
\hline
JRuby & 
\texttt{// All errors to sysread should be SystemCallErrors, but on a closed stream Ruby returns an IOError.  Java throws same exception for all errors so we resort to this hack...} & 
no & 
yes \\
\hline
Columba & 
\texttt{// FIXME r.setPos();} & 
no & 
yes \\

\end{tabular}
\vspace{-20pt}
\label{tab:updated_examples}
\end{table}
 To structure our work,  we investigate the following research questions:


\textbf{RQ1: Can we still reduce the cost of labeling for static warning analysis?} Due to data duplication and features leakage, static warning analysis's data from \citet{tu_frugal21} is problematic. Thanks to Kang et al.\cite{kang_static}, the updated data is utilized here to revalidate FRUGAL's effectiveness against the SOTA for adoptable static warning identification. 

\begin{RQ}{\normalsize{Result:}} 
FRUGAL exceeds the SOTA SVM (EMSE'20 \cite{intrinsic_static} in identifying adoptable static warnings). FRUGAL requires only 2.5\% at median of the train data to be labeled  while using 97.5\% less information than the SOTA SVM. 
\end{RQ} \vspace{-10pt}

\textbf{RQ2: How well does FRUGAL distinguish security bug reports?} From our investigation of various $L\%$ values in this task, FRUGAL’s performance still suffice at $2.5\%$.  FRUGAL outshines the SOTA solution in EMSE'21 in 2/3 metrics. However, the improvement is relatively less impressive in comparison to previously reported in issue close time prediction and static warning analysis due to highly imbalanced data nature and insufficient training data. 

\begin{RQ}{\normalsize{Result:}} 
FRUGAL exceeds the SOTA SWIFT (EMSE'21 \cite{Shu2021HowTB} in distinguishing security bug reports). FRUGAL requires only 2.5\% at median of the train data to be labeled  while using 97.5\% less information than the SOTA SWIFT. 
\end{RQ} \vspace{-10pt}

\textbf{RQ3: Can FRUGAL better mitigate biases in SE ML models with less data?} From our investigation of various $L\%$ values in this task, FRUGAL’s performance still suffice at $2.5\%$.  In fairness management of machine learning software, FRUGAL outshines the SOTA solution in FSE'21 in 7/9 metrics.  This analysis indicates how FRUGAL can ge generalized to non-SE tasks if their data does share similar nature as SE data. \vspace{-5pt}

\begin{RQ}{\vspace{-30pt}\normalsize{Result:}} \vspace{-2pt}
Even with non-SE data, FRUGAL still exceeds the SOTA FAIR\_SMOTE (FSE'21 \cite{fair_smote} in managing fairness in machine learning software). FRUGAL requires only 2.5\% at median of the train data to be labeled when being compared against unsupervised learning while using 97.5\% less information than both SOTA methods.
\end{RQ} \vspace{-16pt}

\textbf{RQ4: What labeling method would we recommend for SE data?}
Summarizing all the above results, here we contrast the Frugal strong-heuristic
with   weak-heuristic methods from standard AI
(co-training, self-training, label propagation, and label spreading and all the methods
described in  \S\ref{standardssl}). We will conclude that:

\begin{RQ}{\normalsize{Result:}} \vspace{-2pt}
FRUGAL outperforms all standard semi-supervised learning method from the machine learning community across four software analytics tasks. Moreover, this ineffectiveness is hypothetically due to the (1) highly imbalanced data nature  that is not often observed in standard ML tasks; and (2) these standard SSL methods does not tap into that SE domain knowledge that FRUGAL has access to. 
\end{RQ} \vspace{-16pt}

\subsection{Contributions}
This work makes several contributions: 
\be
\item This work demonstrates the benefits of Semi-Supervised Learning for Software Analytics as applied to  software fairness, security bug reports, static warning analysis, and issue close time. 
\item
In  the initial conference version
of this paper~\cite{tu2021frugal},
some of the data  (for the static warning analysis) was recently challenged (in a forthcoming ICSE'22 paper) due to data quality issues. Here, we take the fixed data resulting from that refutation
study~\cite{kang2022detecting} and repeated all the prior analysis. 
\item We extend FRUGAL's effectiveness from SE data to non-SE data (i.e., software fairness). Specifically, FRUGAL is still able to perform well when non-SE data's intrinsic dimensionality is low.
\item Our work incorporates FRUGAL to software fairness and security bug reports with more than 50 datasets, ten times more than the previous study. FRUGAL still outperforms both tasks while using 97.5\% less data. 
\item Our and previous work conclusions are further generalized and strengthened by contrasting against the standard SSL methods from the machine learning community. 
\item To better support other researchers our scripts and data are on-line at
https://github.com/HuyTu7/FRUGAL.
\ee

\subsection{Connection to Prior Work}
 This paper studies four domains:
\begin{enumerate}
\item
Distinguishing security-related  bug reports;
\item
Mitigating bias in decision making;
\item
Predicting issue close time; and 
\item
Reducing false alarms in   static code warnings.
\end{enumerate}
The analysis of items \#1,\#2  has not appeared before.
Preliminary results
on \#3 and \#4 appeared in an ASE'21 paper~\cite{tu2021frugal}.
A forthcoming ICSE'22~\cite{kang_static} paper found that the data used
for the analysis of \#4 has numerous errors. 
Hence, this paper repeats the entire analysis of
\#4, but this time with the corrected data. 
\subsection{Structure of this Paper}
The rest of this paper is structured as follows. \textit{Section 2} discusses the motivation, background and related works. \textit{Section 3} describes our methodology. \textit{Section 4} focuses on our experimental design, while \textit{Section 5} analyzes the results. \textit{Section 6} and \textit{7} discuss our short-comings and directions for future work, respectively.
 
\section{Motivation and Background}
\subsection{Standard versus Specialized Methodology for Software Analytics?}

The SE research have limited exposures to semi-supervised learning. For instance, a lot of recent semi-supervised learning works~\cite{yu2019improving, Yu19, tu_debtfree21, tu_frugal21, tu2020better} within the SE community are mostly classified as one class of SSL, called self-training (such as ~\citet{self_training}). Other classes can also include label propagation (such as ~\citet{Zhu02learningfrom}), majority voting (such as ~\citet{co_training}), and label spreading (such as ~\citet{label_spread}). Most readers to the AI literature would come away with a common picture on how to perform semi-supervised learning based on these four standard semi-supervised learners. 

So, does it suffice to apply standard semi-supervised learning techniques to the SE community? Hindle et al.~\cite{hindle2012naturalness} claimed that software programs are written by real people, which are mostly simple and rather repetitive. Hence, 
there are predictable statistical properties embedded within them that can be learned through statistical language models and leveraged for software engineering tasks. Consequently, they believed that standard techniques from the Natural Language Processing community can be applied to software analytics as well (e.g., code search, summarization, classification, etc). For instance, software researchers~\cite{Attention_SATD} have adopted and applied the SOTA attention-based neural networks~\cite{attention_neurips} from the NLP community to learn source code comments to identify self-admitted technical debts.  

However, in this paper, we hypothesize that it is better to not adopt off-the-shelf technique blindly, and instead, to transform the technique specific to the software analytics community. Russell et al.\cite{AI_4thedition} coined strong versus weak heuristic where weak heuristic relies on general algorithms while strong heuristic leverages on specific domain knowledge. There are many examples of strong versus weak heuristic in software engineering: 
\bi
\item To plan for defect reduction, instead of using plausible changes (i.e., changes with some precedence in the prior releases), Peng et al.~\cite{9371412} focused the plans to just those attributes which change the most within a project that is specialized to this domain. 
\item Novielli et al.~\cite{MSR_satool} reported that their sentiment analysis tool did better when trained on SE data (e.g., Stack Overflow and Jira), not the off-the-shelf New York Times corpus.
\item In defect prediction research, Fu et al.~\cite{fu16} conducted a literature review of 50 highly cited papers in the past decade and found that 80\% of the papers only applied off-the-shelf methods. Yet, when they tuned their defect predictors to the specific SE data, they did significantly better (i.e., precision changes from 0\% to 60\%).
\item Moreover, Agrawal et al.~\cite{agrawal2019dodge} documented how SE data are not as complex as standard ML data. Hence, they leveraged that to build their own technique by ``DODGE-ing'' (i.e. simply steering way from settings that lead to similar conclusions) which outperformed SOTA works across several software analytics tasks. 
\ei  
For the rest of this paper, we check if 
our strong heuristic (big things tend to have more problems)
lets us reason ``faster'' about software projects.
Here, by ``faster'' we do not mean ``less CPU time''
but rather ``less manual work labelling instances''.

\subsection{Static Code Warning}

This paper explores SSL approach with SE domain knowledge as a strong approach across four domains. Our first domain is
static code warnings. The goal of this domain is to identify such a static code warning adoptable or unadoptable. The other domains are discussed
in the next few sections.

\subsubsection{Background}

It has been 15 years since Findbugs~\cite{4602670} was introduced 15 years ago as an automatic static analysis tool (ASAT) to detect bugs in Java programs. These ASATs detect potential static code defects in source code or executable files at the stage of software product development by matching code against bug patterns. These issues include common programming errors, code styling, in-line comments, style violations, and questionable coding decisions, for instance, patterns of
code that may dereference a null pointer. Detecting this early in the project life cycle would better save projects from technical debts. 

However, the warnings generated from these tools
are not guaranteed to be real bugs. Many developers do not perceive the warnings by
ASATs to be relevant due to the high incidence of effective false
alarms~\cite{10.5555/2486788.2486877}. Such warnings are considered as ``unadoptable'' since developers
just ignored them. Between 35\% and 91\% of the warnings generated
from static analysis tools are known to be unadoptable. Hence, SE researchers have extensively studied different warnings programmers usually act upon
so the tools can be made more useful by first pruning away the
unadoptable warnings~\cite{kang_static, wang2018there, incre_static, intrinsic_static}. They also have proposed techniques to further reduce the false alarms and focused on the identifying the actionable warnings. 

\begin{table}[!t]
\small
\vspace{-5pt}
\caption{Distribution summary of \citet{wang2018there}'s problematic data and \citet{kang_static}'s updated data. The gray cells are median values for the corresponding columns. This demonstrates 90\% from previous studies~\cite{wang2018there, incre_static, intrinsic_static, tu_frugal21} are unusable and previous work's conclusions should be assessed again.}
\vspace{-13pt}
\small
\begin{center}
\begin{tabular}{@{}rrrrrrr@{}}
\midrule
\rowcolor[HTML]{7DFFEA}  & \textbf{Dataset} &  \textbf{Features}& 
 \begin{tabular}[c]{@{}c@{}}\textbf{Wang et al.'s\cite{wang2018there}} \\ \textbf{Data}\end{tabular} & \begin{tabular}[c]{@{}c@{}}\textbf{Kang et al.'s\cite{kang_static}} \\ \textbf{Data}\end{tabular} & 
 
 \\ \midrule
 
 & commons &39& 1511 & \cellcolor{gray!20}22  \\ 
 & phoenix &44&  4524 & 0  \\
 & mvn (maven) & 47 & 1631 & 4  \\
 & jmeter & 49 & 1217 & 18  \\
 & cass (cassandra) &\cellcolor{gray!20}55& 5185 & 17  \\
 & ant &56& 2344 & 35  \\
 & lucence &57& 6684 & 30  \\
 
 & derby &58& 4986 & 458  \\
 
 & tomcat &60& \cellcolor{gray!20}2876 & 184  \\
 \bottomrule
\end{tabular}
\end{center}
\vspace{-17pt}

\label{table:distribution}
\end{table}

Recently, Wang et al.~\cite{wang2018there} completed
a systematic evaluation of the features that have been proposed in
the literature, and identified 23 ``Golden Features''
that seemed more useful for recognizing
useful static code warnings.
Using these features, subsequent studies \cite{intrinsic_static, incre_static} showed that
any off-the-shelf standard machine learning techniques, e.g. SVM, can perform effectively,
and that the use of a small number of training instances can train
effective models. These studies have reported the performances of up to 96\% Recall,
98\% Precision, and 99.5\% AUC. \citet{intrinsic_static} identified the nature of these strong results within the task of
detecting actionable warnings as ``intrinsically easy''. This is essentially the domain knowledge within this task specifically and SE data generally that our FRUGAL would leverage on as a strong heuristic. However, Kang et al.~\cite{kang_static} found evidences that the data utilized by prior studies and Tu et al.'s FRUGAL have data leakage and data duplication. Data leakage involves how the ground-truth labels have leaked into the features that compute the proportion of actionable warnings. Data duplication is simply many train data instances also appear in the testing dataset. These issues threaten the conclusion validity of prior studies including half of the data in Tu et al.\cite{tu_frugal21}'s study to propose FRUGAL. This reaffirms the importance of data quality and effort within SE community to motivate this study.

\subsubsection{Data and Algorithm}

The data for this paper originally came from a recent study by Kang et al.~\cite{kang_static}. Table~\ref{table:distribution} summarizes the data distribution after removing the data leakages and data duplication from Wang et al.~\cite{wang2018there}. The original 30,000 data instances in Wang et al.~\cite{wang2018there}'s study were reduced to 768. Moreover, the percentage of actionable warnings within the data on average was 15\% in \citet{wang2018there}'s study but 39.7\% in \citet{kang_static}'s study. This shows that at least 90\% of the previous data is unusable and previous work should be assessed again. However, regarding the nature of static warnings, Wang et al.~\cite{wang2018there} completed
a systematic evaluation of the features that have been proposed in
the literature, and identified 23 ``Golden Features'', i.e., the
most important features for detecting actionable warnings. Their data is problematic but this is the most exhaustive research about static warning characteristics yet published.
As shown in Table~\ref{table:variables},
the  ``golden set'' features are the independent variables fall into seven categories. To assign dependent labels, we applied the methods of
Liang et al.~\cite{liang2010automatic}.
They defined a specific warning as adoptable if it is closed after the later revision interval. However, seven out of nine projects have less than 40 data instances which is below than recommended (at least 75 per class~\cite{samplesize}). Consequently, we combine all the projects' data together to have a sufficient amount of data for training and testing. 

\begin{table}[!t]
\footnotesize
\vspace{-5pt}
\caption{Categories of Wang et al.~\cite{wang2018there}'s 23 Golden Features which was originally 95 features spanning across 8 categories. }
\vspace{-5pt}
\tabcolsep=0.2cm
\begin{center}
\begin{tabular}{ll}
\hline
\rowcolor[HTML]{7DFFEA} \textbf{Category}  &   \textbf{Features} \\ \hline
\multirow{1}{*}{\rotatebox[origin=c]{0}{\parbox[c]{1cm}{ Warning  \\ Combination}}}  & \begin{tabular}[c]{@{}l@{}} warning context in method, file\\ warning context for warning type;\\ defect likelihood for warning pattern;\\ discretization of defect likelihood; \\ average lifetime for warning type;\end{tabular} \\ \hline
\multirow{1}{*}{\rotatebox[origin=c]{0}{\parbox[c]{1cm}{ Code  \\ characteristics}}}   & \begin{tabular}[c]{@{}l@{}} comment-code ratio;\\ method, file depth;\\ methods in file\\package;\end{tabular} \\ \hline
\multirow{1}{*}{\rotatebox[origin=c]{0}{\parbox[c]{0cm}{ Warning\\ characteristics}}}  & \begin{tabular}[c]{@{}l@{}}warning pattern; \\ 
type, priority;  \\  package;\end{tabular} \\ \hline
\multirow{1}{*}{\rotatebox[origin=c]{0}{\parbox[c]{2cm}{ File  history}}}    & \begin{tabular}[c]{@{}l@{}} file age; file creation;\\ developers;\end{tabular} \\ \hline
\multirow{1}{*}{\rotatebox[origin=c]{0}{\parbox[c]{2cm}{ Code  analysis}}}   & \begin{tabular}[c]{@{}l@{}}parameter signature,\\ method visibility\end{tabular} \\ \hline
\multirow{1}{*}{\rotatebox[origin=c]{0}{\parbox[c]{2cm}{ Code  history}}}   
 & \begin{tabular}[c]{@{}l@{}}added LOC in file in the last 25 revisions;\\ added LOC in package in the past 3 months;\end{tabular} \\ \hline
\multirow{1}{*}{\rotatebox[origin=c]{0}{\parbox[c]{2.5cm}{ Warning  history}}}   & \begin{tabular}[c]{@{}l@{}}warning lifetime by revision;\end{tabular} \\ \hline

\end{tabular}
\end{center}
\label{table:variables} 
\vspace{-10pt}
\end{table}

\subsection{Security Bug Reports Categorization}

Our second domain is security bug reports. In this domain, the goal is to categorize such a bug report is a security one or not. 

\subsubsection{Background}

Security bugs have been a pressing concern within software analytics. A NIST's report commented that \textit{``current systems perform increasingly vital tasks and are widely known to possess vulnerabilities''} \cite{black2016dramatically} where \textit{vulnerability} is a weakness in the computational logic (e.g., code) that, when exploited, results in a negative impact on confidentiality, integrity, or availability \cite{lewiscommon}. Daily news reports increasingly sophisticated security breaches. As seen in those reports, a single vulnerability can have devastating effects. For example, the WannaCry ransomware attack \cite{WannaCry} crippled British medical emergency rooms, delaying medical procedures for many patients. A data breach of Equifax caused many personal information lost, as many as 143 million Americans -- or nearly half the country -- to be compromised \cite{Equifax}. 

\begin{table*}[!t]
\caption {Different filters used in FARSEC.}
\centering
\begin{tabular}{l|l}
\hline
\rowcolor[HTML]{7DFFEA} 
\multicolumn{1}{c|}{\cellcolor[HTML]{7DFFEA}\textbf{Filter}} & \multicolumn{1}{c}{\cellcolor[HTML]{7DFFEA}\textbf{Description}} \\ \hline
farsecsq & \begin{tabular}[c]{@{}l@{}}Apply the Jalali et al.~\cite{jalali2008optimizing} support function\\ to the frequency of words found in SBRs\end{tabular} \\ \hline
farsectwo & \begin{tabular}[c]{@{}l@{}}Apply the Graham version~\cite{graham2004hackers} of multiplying \\ the frequency by two.\end{tabular} \\ \hline
farsec & Apply no support function. \\ \hline
clni & \begin{tabular}[c]{@{}l@{}} Apply Closet List Noise Identification \\ (CLNI~\cite{kim2011dealing}) filter to non-filtered data.\end{tabular} \\ \hline
clnifarsec & Apply CLNI filter to farsec filtered data. \\ \hline
clnifarsecsq & Apply CLNI filter to farsecsq filtered data. \\ \hline
clnifarsectwo & Apply CLNI filter to farsectwo filtered data. \\ \hline
\end{tabular}
\label{tbl:farsecFilter}
\end{table*}

Developers capture and document software bugs and issues into bug reports which are submitted to bug tracking systems. For example, the Mozilla bug database maintains more than 670,000 bug reports with 135 new bug reports added each day~\cite{chen2013r2fix}. Submitted bug reports are explicitly labeled as a security bug report (SBR) or non-security bug report (NSBR). Within such bug tracking systems, Peters et al.~\cite{peters2018text} warn that it is important to correctly identify security bug reports and distinguish them from other non-security bug reports. They note that software vendors ask that security bug reports should be reported directly and privately to their own engineers. These engineers then assess the bug reports and, when necessary, offer a security patch. The security bug, and its associated patch, can then be documented and disclosed via public bug tracking systems. This approach maximizes the probability that a patch is widely available before hackers exploit a vulnerability. Sometimes, bug reporters lack the security domain knowledge~\cite{gegick2010identifying} to know when their bug is a normal bug (which can be safely disclosed) or when that bug is a security bug (that needs to be handled more discretely). Hence, lamentably, security bugs are often publicly disclosed before they can be patched~\cite{wijayasekara2014vulnerability}.



\subsubsection{Data and Algorithms}

\begin{table*}[!t]
\scriptsize
\caption {Imbalanced characteristic of bug report data sets from FARSEC~\cite{peters2018text}.}
\setlength\tabcolsep{4pt}
\centering
\begin{tabular}{c|l|c|r|c|c|r|c}
\hline
\rowcolor[HTML]{7DFFEA} 
\cellcolor[HTML]{7DFFEA} & \multicolumn{1}{c|}{\cellcolor[HTML]{7DFFEA}} & \multicolumn{3}{c|}{\cellcolor[HTML]{7DFFEA}\textbf{Training}} & \multicolumn{3}{c}{\cellcolor[HTML]{7DFFEA}\textbf{Testing}} \\ \cline{3-8} 
\rowcolor[HTML]{7DFFEA} 
\multirow{-2}{*}{\cellcolor[HTML]{7DFFEA}\textbf{Project}} & \multicolumn{1}{c|}{\multirow{-2}{*}{\cellcolor[HTML]{7DFFEA}\textbf{Filter}}} & \textbf{\#SBRs} & \multicolumn{1}{c|}{\cellcolor[HTML]{7DFFEA}\textbf{\#BRs}} & \textbf{SBRs(\%)} & \textbf{\#SBRs} & \multicolumn{1}{c|}{\cellcolor[HTML]{7DFFEA}\textbf{\#BRs}} & \textbf{SBRs(\%)} \\ \hline
 & train &  & 20,970 & 0.37 &  &  &  \\ 
 & farsecsq &  & 14,219 & 0.54 &  &  &  \\ 
 & farsectwo &  & 20,968 & 0.37 &  &  &  \\ 
 & farsec &  & 20,969 & 0.37 &  &  &  \\ 
 & clni &  & 20,154 & 0.38 &  &  &  \\ 
 & clnifarsecsq &  & 13,705 & 0.56 &  &  &  \\ 
 & clnifarsectwo &  & 20,152 & 0.38 &  &  &  \\ 
\multirow{-8}{*}{Chromium} & clnifarsec & \multirow{-8}{*}{77} & 20,153 & 0.38 & \multirow{-8}{*}{115} & \multirow{-8}{*}{20,970} & \multirow{-8}{*}{0.55} \\ \hline
 & train &  & 500 & 0.80 &  &  &  \\
 & farsecsq &  & 136 & 2.94 &  &  &  \\ 
 & farsectwo &  & 143 & 2.80 &  &  &  \\ 
 & farsec &  & 302 & 1.32 &  &  &  \\ 
 & clni &  & 392 & 1.02 &  &  &  \\ 
 & clnifarsecsq &  & 46 & 8.70 &  &  &  \\ 
 & clnifarsectwo &  & 49 & 8.16 &  &  &  \\ 
\multirow{-8}{*}{Wicket} & clnifarsec & \multirow{-8}{*}{4} & 196 & 2.04 & \multirow{-8}{*}{6} & \multirow{-8}{*}{500} & \multirow{-8}{*}{1.20} \\ \hline
 & train &  & 500 & 4.40 &  &  &  \\ 
 & farsecsq &  & 149 & 14.77 &  &  &  \\ 
 & farsectwo &  & 260 & 8.46 &  &  &  \\ 
 & farsec &  & 462 & 4.76 &  &  &  \\ 
 & clni &  & 409 & 5.38 &  &  &  \\ 
 & clnifarsecsq &  & 76 & 28.95 &  &  &  \\ 
 & clnifarsectwo &  & 181 & 12.15 &  &  &  \\ 
\multirow{-8}{*}{Ambari} & clnifarsec & \multirow{-8}{*}{22} & 376 & 5.85 & \multirow{-8}{*}{7} & \multirow{-8}{*}{500} & \multirow{-8}{*}{1.40} \\ \hline
 & train &  & 500 & 2.80 &  &  &  \\ 
 & farsecsq &  & 116 & 12.07 &  &  &  \\ 
 & farsectwo &  & 203 & 6.90 &  &  &  \\ 
 & farsec &  & 470 & 2.98 &  &  &  \\ 
 & clni &  & 440 & 3.18 &  &  &  \\ 
 & clnifarsecsq &  & 71 & 19.72 &  &  &  \\ 
 & clnifarsectwo &  & 151 & 9.27 &  &  &  \\ 
\multirow{-8}{*}{Camel} & clnifarsec & \multirow{-8}{*}{14} & 410 & 3.41 & \multirow{-8}{*}{18} & \multirow{-8}{*}{500} & \multirow{-8}{*}{3.60} \\ \hline
 & train &  & 500 & 9.20 &  &  &  \\ 
 & farsecsq &  & 57 & 80.70 &  &  &  \\ 
 & farsectwo &  & 185 & 24.86 &  &  &  \\ 
 & farsec &  & 489 & 9.41 &  &  &  \\ 
 & clni &  & 446 & 10.31 &  &  &  \\ 
 & clnifarsecsq &  & 48 & 95.83 &  &  &  \\ 
 & clnifarsectwo &  & 168 & 27.38 &  &  &  \\ 
\multirow{-8}{*}{Derby} & clnifarsec & \multirow{-8}{*}{46} & 435 & 10.57 & \multirow{-8}{*}{42} & \multirow{-8}{*}{500} & \multirow{-8}{*}{8.40} \\ \hline
\end{tabular}
\label{tbl:farsecDataset}
\end{table*}

Peters et al. recently proposed FARSEC~\cite{peters2018text} work where they reported more success after focusing on a particular problem within the security domain. FARSEC is a technique that adds an irrelevancy pruning step to data mining in building security bug prediction models. \tbl{farsecFilter} lists the pruners explored in the FARSEC research. The purpose of filtering in FARSEC is to remove non-security bug reports with security related keywords. To achieve this goal, FARSEC applied an algorithm that firstly calculated the probability of the keywords appearing in security bug report and non-security bug report, and then calculated the score of the keywords.

Inspired by previous works~\cite{graham2004hackers,jalali2008optimizing}, several tricks were also introduced in FARSEC to reduce false positives. These filters include: 
\bi
\item {\em farsectwo}: multiplying the frequency of non-security bug reports by two, i.e., to achieve a good bias.
\item {\em farsecsq}: squaring the numerator of the support function to improve heuristic ranking of low frequency evidence. 
\item {\em CLNI} (Closet List Noise Identification)~\cite{kim2011dealing}: a noise detection algorithm and removal that is based on Euclidean distance. 
\ei

One of the common issues with imbalanced data prediction is the large number of false positives in the prediction results. This matters because it means the model's performance is not sufficient and especially extra effort is required from developers to check those false positives. One possible solution is generating a list of ranked bug reports, as done via FARSEC. This method takes two steps:

\begin{enumerate}
    \item For a filter $f$, the ranked prediction results are selected from non-filtered data or data with filters other than $f$ which has less number of predicted security bug reports than filter $f$. 
    \item If the first step is not applicable, the chronological order is utilized. The predicted security bug reports are then prioritized as close to the top of the list than non-security bug reports.
\end{enumerate}

 Table~\ref{tbl:farsecDataset} elaborates on the characteristics of the FARSEC datasets. Our experiments reproduce and improve the FARSEC results from the SOTA SWIFT~\cite{Shu2021HowTB} using the same datasets. As we see from the table, one unique feature of the data set is the rarity of the target class. The ``SBRs\%'' column in both training and testing data set indicates that security bug reports make up a very small percentage of the total number of bug reports in projects like Chromium.

\subsection{Software Fairness}

Our third domain is software fairness. Similarly, the goal is still binary classification but not only to improve the performance metrics but also the associated fairness metrics (to be defined in \S2.5.1). Moreover, each dataset here has their own class label, e.g., Home Credit dataset's goal is to approve or reject a loan application from an individual or Compas's goal is to predict re-offend/no re-offend based on criminal history of defendants.

\subsubsection{Background}

Software is as important as it has ever been in our current society. Many processes that were once performed by humans have been automated by software, as the advances in machine learning, automation, big data, and artificial intelligence have granted software the capacity to make decisions. Every day, these systems take these decisions in our stead, ranging from health and diagnostics, bank loaning, automated vehicles, recommendation systems, and even criminal justice systems. Unfortunately, there are too many recent examples where models learned from AI are demonstrably discriminatory towards certain social sub-groups.  Examples of this include:

\begin{itemize}
\item Translation and image search that exhibit gender stereotypes~\cite{caliskan2017semantics}; 
\item Risk-assessment score computation used in assigning bond amounts and sentencing in the US criminal justice system, exhibits racial bias~\cite{angwin2016machine}; 
\item Facial detection and recognition tools’ accuracy depends on demographic information, such as race and gender~\cite{klare2012face};
\item Amazon’s software for automatically deciding where to offer same day delivery excluded areas by socio-economic indicators~\cite{ingold2016amazon};
\item Online search engines have been more likely to display ads related to arrest records with searches for traditionally-minority names~\cite{sweeney2013discrimination};
\item Orbitz.com has steered Mac users to more expensive hotels~\cite{mattioli2012orbitz};
\item Tools with natural English inputs parse English written by white people more accurately than that written by people of other races~\cite{blodgett2017racial};
\item Alphabet's YouTube auto-captioning service’s accuracy is
higher for male voices than female ones~\cite{tatman2017gender}. 
\end{itemize}

As it stands, software fairness testing is an under-explored discipline, often omitted from the top-level decisions when developing these decision-making systems~\cite{galhotra2017fairness}.
This represents
a hole in current research since,
recently, the requirements for fairer AI
have become more common.
The European Union
and Microsoft and the IEEE have all released white papers discussing fair and ethical AI~\cite{IEEEethics,EU,MicrosoftEthics}.
While these  documents differ in the details,  they all agree that  ethical AI should must be ``FAT''; i.e.  fair, accountable and transparent. 

Recently, the software engineering and machine learning community have become interested in the problem of fairness.  ICSE and ASE conducted separate workshops for software fairness \cite{FAIRWARE,EXPLAIN}.  ACM and IEEE have started
separate conferences like FAccT~\cite{faact} and FILA~\cite{FILA} for fairness of ML
models. Big software industries have started taking this fairness problem seriously, e.g., IBM~\cite{AIF360}, Microsoft~\cite{FATE}, Facebook/Meta~\cite{Facebook}, etc. 

\subsubsection{Data and Algorithms}

 Table \ref{tbl:fairness_data} contains the datasets  used in this study. All of them are binary classification problems where the target class has only two values.
\bi
\item
A class label is called a \textit{favorable label} if it gives an advantage to the receiver such as receiving a loan, being hired for a job.
\item
A \textit{protected attribute} is an attribute that divides the whole population into two groups (privileged \& unprivileged) that have differences in terms of   receiving benefits. 
\ei
  Every dataset in Table~\ref{tbl:fairness_data} has one or at most two protected attributes. For example, in case of credit card application datasets, based on protected attribute ``sex'', ``male'' is privileged and ``female'' is unprivileged; in case of Medical Expenditure Panel Survey (MEPS) datasets, based on protected attribute ``race'', ``white'' is privileged and ``POC'' (i.e., person of color) is unprivileged.
\bi
\item
\textit{Group fairness} is the goal that based on the protected attribute, privileged and unprivileged groups will be treated similarly.
\item
\textit{Individual fairness} means similar outcomes go to similar individuals. 
\ei

To our knowledge, there are only two frameworks Fairway~\cite{fairway20} and FAIR\_SMOTE~\cite{fair_smote} that attempted to mitigate bias in SE ML models. Both of them are supervised methods that require a lot of labeled training data. Fairway~\cite{fairway20} focused on removing the biased labels while FAIR\_SMOTE~\cite{fair_smote} balanced the data to achieve fairer model and better performance. Since FAIR\_SMOTE outperformed Fairway, we will employ FAIR\_SMOTE as the SOTA work for this software analytics task.

\begin{table*}[!t]
\setlength\tabcolsep{1pt}
\renewcommand{\arraystretch}{1.1}
\centering
\caption{Details of the software fairness datasets used in this study.}
\label{tbl:fairness_data}
\scriptsize
\begin{tabular}{ccccccc}
\toprule
\rowcolor[HTML]{7DFFEA}  
 &  &  & \multicolumn{2}{c}{\cellcolor[HTML]{7DFFEA}\textbf{Protected Attribute}}                                                                   & \multicolumn{2}{c}{\cellcolor[HTML]{7DFFEA}\textbf{Class Label}}                                                              \\ \cline{4-7}
\rowcolor[HTML]{7DFFEA} 
\multirow{-2}{*}{\textbf{Dataset}}  & \multirow{-2}{*}{\textbf{\#Rows}}       & \multirow{-2}{*}{\textbf{\#Columns}}       & \textbf{Privileged}   & \textbf{Unprivileged}      & \textbf{Favorable}       & \textbf{Unfavorable} \\ \toprule

\begin{tabular}[c]{@{}c@{}}Adult \\ Census\\ Income\end{tabular} & 48,842          & 14              & \begin{tabular}[c]{@{}c@{}}Sex: Male\\ Race: White\end{tabular}     & \begin{tabular}[c]{@{}c@{}}Sex: Female\\ Race: POC\end{tabular}     & \begin{tabular}[c]{@{}c@{}}High \\ Income\end{tabular}         & \begin{tabular}[c]{@{}c@{}}Low \\  Income\end{tabular}         \\ \midrule
Compas                                                           & 7,214           & 28              & \begin{tabular}[c]{@{}c@{}}Sex: Female\\ Race: White\end{tabular} & \begin{tabular}[c]{@{}c@{}}Sex: Male\\Race: POC\end{tabular} & \begin{tabular}[c]{@{}c@{}}Did not \\ reoffend\end{tabular}    & Reoffended                                                    \\ \midrule
\begin{tabular}[c]{@{}c@{}}German \\ Credit\end{tabular}         & 1,000           & 20              & Sex-Male                                                          & Sex-Female                                                              & Good Credit                                                    & \begin{tabular}[c]{@{}c@{}}Bad \\ Credit\end{tabular}         \\ \midrule
\begin{tabular}[c]{@{}c@{}}Default \\ Credit\end{tabular}        & 30,000          & 23              & Sex-Male                                                          & Sex-Female                                                              & \begin{tabular}[c]{@{}c@{}}Default \\ Payment-Yes\end{tabular} & \begin{tabular}[c]{@{}c@{}}Default \\ Payment-No\end{tabular} \\ \midrule
\begin{tabular}[c]{@{}c@{}}Heart \\ Health\end{tabular}          & 297             & 14              & Age-Young                                                         & Age-Old                                                                 & \begin{tabular}[c]{@{}c@{}}Not \\ Disease\end{tabular}         & Disease                                                       \\ \midrule
\begin{tabular}[c]{@{}c@{}}Bank \\ Marketing\end{tabular}        & 45,211          & 16              & Age-Old                                                           & Age-Young                                                               & \begin{tabular}[c]{@{}c@{}}Term \\ Deposit - Yes\end{tabular}  & \begin{tabular}[c]{@{}c@{}}Term \\ Deposit - No\end{tabular}  \\ \midrule
\begin{tabular}[c]{@{}c@{}}Home \\ Credit\end{tabular}           & 3,075,11        & 240             & Sex: Male & Sex: Female & Approved   & Rejected                                                      \\ \midrule
\begin{tabular}[c]{@{}c@{}}Student \\ Performance\end{tabular}   & 1,044 & 33 & Sex: Male & Sex: Female  & \begin{tabular}[c]{@{}c@{}}Good \\ Grade\end{tabular}          & \begin{tabular}[c]{@{}c@{}}Bad \\ Grade\end{tabular}          \\ \midrule
\begin{tabular}[c]{@{}c@{}}MEPS15,\\ MEPS16\end{tabular}         & 35,428          & 1,831           & Race: White                                                        & Race: POC                                                          & \begin{tabular}[c]{@{}c@{}}Good \\ Utilization\end{tabular}    & \begin{tabular}[c]{@{}c@{}}Bad \\ Utilization\end{tabular}    \\ \bottomrule
\end{tabular}
\end{table*}

\subsection{Evaluation}

\subsubsection{Measures of Performance}\label{measures}

Since we wish to compare our approach to prior work, we take the methodological step of adopting the same performance scores as that seen in prior work. Let TP, TN, FP, FN are the true positives, true negatives, false positives, and false negatives (respectively), then \citet{Shu2021HowTB} used recall, false-alarm, and IFA while \citet{fair_smote} used recall, false-alarm, accuracy, precision, F1 for their studies (but see our cautionary note
at the end of this list on precision and F1). We also include AUC and accuracy for validation in RQ3:



\begin{itemize}
    \item \textbf{Recall} = $TP/(TP+FN)$ represents the ability of one algorithm to identify instances of positive class from the given dataset.
    \item \textbf{False Alarms (FAR)} = $TN/(TN+FP)$ measures the instances that are falsely classified by an algorithm as positive which are actually negative. This is an important index used to measure the efficiency of a model.
    \item \textbf{Precision} = $TP/(TP+FP)$ represents the ability of one algorithm to identify instances of positive class among the retrieved positive instances. 
    \item \textbf{F1} = $(2 * Precision * Recall)/(Precision + Recall)$ is the harmonic mean of both precision and recall metrics.
    \item \textbf{Accuracy} = $\mathit{(TP+TN)/(TP+TN+FP+FN)}$ is the percentage of correctly classified samples.
    \item \textbf{AUC} (Area Under the ROC Curve) measures the two-dimensional area under the Receiver Operator Characteristic (ROC) curve~\cite{witten2016data,heckman2011systematic}. It provides an aggregate and overall evaluation of performance across all possible classification thresholds to overall report the discrimination of a classifier~\cite{wang2018there}. 
    \item
    In the effort-aware theme of this paper, we are   interested in the labeling effort to commission new models building which is \textbf{Cost} = $ \frac{|\{\text{human verified comments}\}|}{|\{\text{comments}\}|}$. 
    \item Except for FAR and Cost, for the rest of these metrics (Accuracy, Recall, and AUC), the {\underline{\em higher}} the value,
the 
{\underline{\em better}} the performance.

\end{itemize}
  {\bf Cautionary note:}   \citet{Menzies:2007prec} warns that precision can be misleading for imbalanced data sets like that studied here
(e.g.   Table~\ref{tbl:farsecDataset} reports that for static warning analysis, the median of target class is 15\%).  Hence, while   we do not place much weight on
classifiers that fail on precision or F1.

Outside of the performance metrics, for the software fairness task, we also employ fairness metrics in order to assess how the methods have mitigated the biases within the SE ML models: 

\begin{itemize}
    \item \textbf{Average Odds Difference (AOD)}: Average of difference in False Positive Rates (FPR) and  True Positive Rates (TPR) for unprivileged and privileged groups \cite{IBM}. TPR = TP/(TP + FN),  FPR = FP/(FP + TN), $AOD = {[}(FPR_{U} - FPR_{P}) + (TPR_{U} - TPR_{P}){]} * 0.5$
    \item \textbf{Equal Opportunity Difference (EOD)}:  Difference of True Positive Rates(TPR) for  unprivileged and privileged groups \cite{IBM}. $EOD = TPR_{U} - TPR_{P}$
    \item \textbf{Statistical Parity Difference (SPD)}: Difference between probability  of unprivileged  group (protected attribute PA = 0) gets favorable prediction ($\hat{Y} = 1$) \& probability  of privileged group (protected attribute PA = 1) gets favorable prediction ($\hat{Y} = 1$) \cite{10.1007/s10618-010-0190-x}. $SPD = P[\hat{Y}=1|PA=0] - P[\hat{Y}=1|PA=1]$
    \item \textbf{Disparate Impact (DI)}: Similar to SPD but instead of the difference of probabilities,  the ratio is measured \cite{feldman2015certifying}. $DI = P[\hat{Y}=1|PA=0]/P[\hat{Y}=1|PA=1]$
\end{itemize}

\subsubsection{Statistical Analysis}

With the deterministic nature, we employed Cohen's $d$ effect size test to determine which results are similar by calculating $medium\_{step2}$ across Recall, False Alarm, AUC, Accuracy, and cost. As to what $d$ to use for this analysis, we take the advice of a widely accepted Sawilowsky et al.'s work~\cite{Sawilowsky2009NewES}. That paper asserts that ``small'' and ``medium'' effects can be measured using $d = 0.2$ and $d = 0.5$ (respectively). Splitting the difference, we will analyze this data looking for differences larger than $d = (0.5 + 0.2)/2 = 0.35$: \vspace{-10pt}


\begin{figure*}[!t]
 
\centering \includegraphics[width=0.95\linewidth]{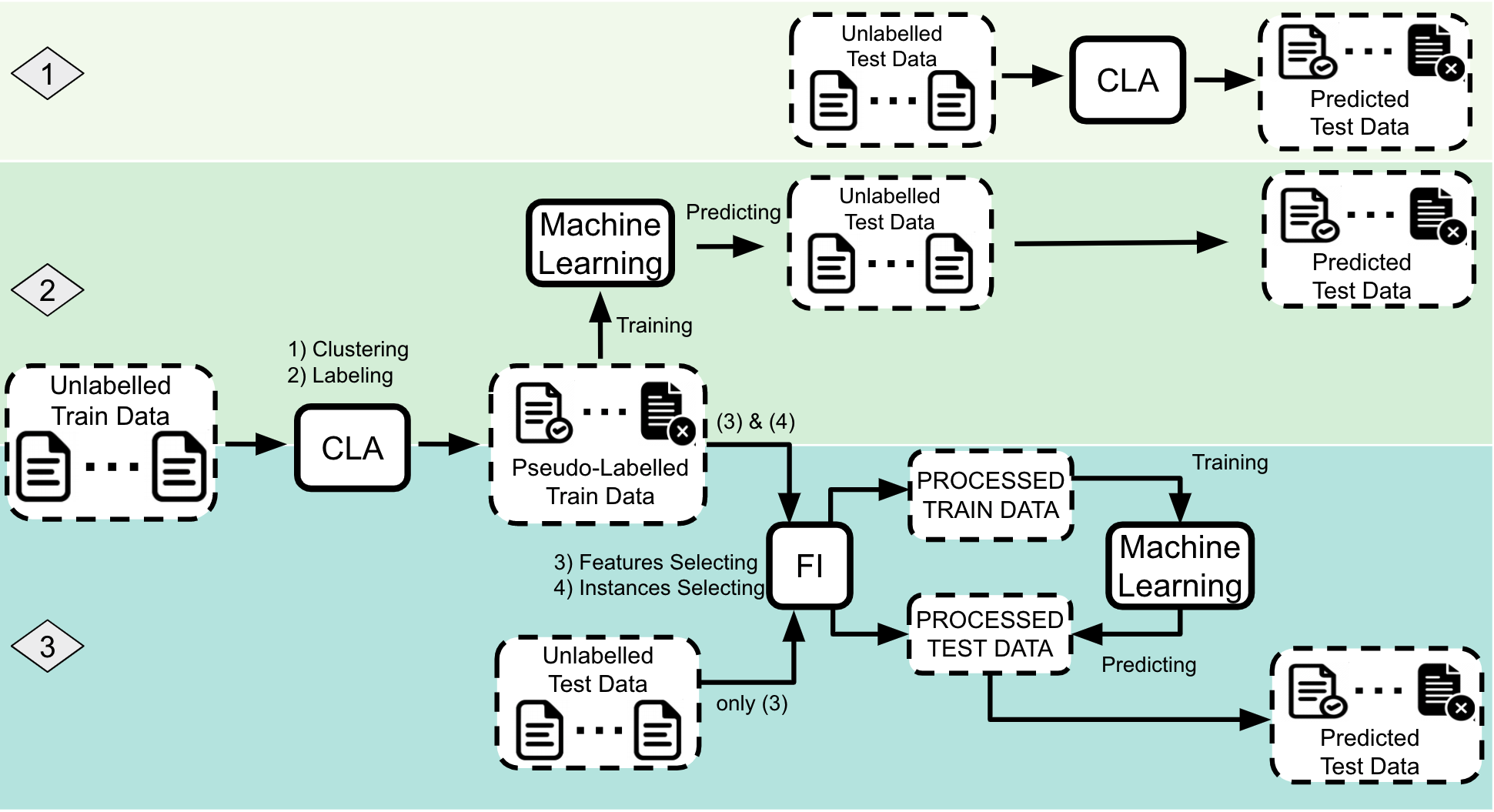}
\vspace{-10pt}
\caption{The FRUGAL system, adopted from Tu et al.\cite{tu_frugal21}.}
 
\label{fig:cla}
\end{figure*} 

\section{Methodology}

\subsection{General Framework}

Tu et al.'s FRUGAL~\cite{tu_frugal21} 
extended Nam et al.'s unsupervised learner, called CLA~\cite{nam2015clami}, by incorporating several variants of CLA, redesigning to be as semi-supervised learning method, and tuning. Nam et al.'s CLA is the SOTA unsupervised learner for defect prediction, which is also confirmed by \citet{unsup_review}'s large-scale study. As shown in Figure~\ref{fig:cla}, CLA consists of three modes: CLA, CLA+ML, and CLAFI+ML. This study shall adopt and extend CLA with tuning in the grid search manner of (1) three modes of CLA while varying (2) the $C\%$ percentile parameter. Simply, FRUGAL finds the best combination of unsupervised learners = \{CLA, CLA+ML, CLAFI+ML\} and $C = \{5\%$ to $95\%$ increments by $5\%\}$. The author only proposed CLA and CLAFI+ML but CLA+ML is a natural medium that can be useful during the tuning process. 

We   explain the details of our approach in   \S3.1, \S3.2, and \S3.3.

\subsection{CLA} \label{sec:cla}

In the SOTA comparative study of unsupervised models in defect prediction, 
CLA starts with two steps of (1) \underline{\textbf{C}}lustering the instances and (2) \underline{\textbf{LA}}belling those instances accordingly to the cluster. In the setting with no train data available, we can label or predict all new/test instances, as shown in the \colorbox[HTML]{f0f9e8}{first block} of Figure~\ref{fig:cla}. 

\noindent \underline{\textbf{Clustering}}: 
\be
\item Find the median of feature $F_1, F_2,..., F_n$ ($percentile(F_i, C)$) where $C = 50\%$ across the whole dataset.
\item For each data instance $X_i$, go through each feature value of the respective data instance to count the time when the feature $F_i > percentile(F_i, C)$ as $K_i$.
\ee

\noindent \underline{\textbf{Labelling}}: label the instance $X_i$ as the positive class if $K_i > median(K)$, else label it as the negative class.

The intuition of such method is based on the defect proneness tendency that is often found in defect prediction research, that is \textit{the higher complexity is associated with the proneness of the defects}  \cite{nam2015clami}. Simply, there is a tendency where the problematic instance's feature values are higher than the non-problematic ones. This tendency and CLA's/CLAFI+ML's effectivenesses are confirmed via the recent literature and comparative study of 40 unsupervised models in defect prediction across 27 datasets and three types of features by \citet{unsup_review}. They found CLA's/CLAFI+ML's performances are superior to other unsupervised methods while similar to supervised learning approaches. Therefore, this study investigated and found that the hypothesized tendency is also applicable in issues close time prediction and adoptable static code warning identification data but not with $C$ at the median ($C = 50\%$). This opens opportunities for hyperparameter tuning. 

\subsection{CLA + ML} \label{sec:cla+ml}

If there is an abundant train data in the wild but without labels, CLA can pseudo-label the train data before applying any machine learner in the ``supervised'' manner (as shown in the \colorbox[HTML]{bae4bc}{second block} in Figure~\ref{fig:cla}). For this step, we take \citet{nam2015clami}'s advice to incorporate Random Forest~\cite{Breiman2001} (RF, described in \S4.5.1), an ensemble of tree learners method, as the machine learner of choice.

\subsection{CLAFI + ML} \label{sec:clafi+ml}

CLAFI is an extension of CLA which is a fullstack framework that also include (3) \underline{\textbf{F}}eatures selection and (4) \underline{\textbf{I}}nstances selection. The setting is similar to CLA+ML, as shown in the \colorbox[HTML]{7bccc4}{third block} of Figure~\ref{fig:cla}, the pseudo-labelled train data (from CLA) and unlabelled test data will be processed with \underline{\textbf{FI}} and \underline{\textbf{F}} respectively. Finally, the machine learner can train the processed pseudo-labelled train data and then predict on the processed test data.   


\noindent \textbf{\underline{Feature Selection}}: Calculate the violation score per feature, called metric in the original proposal of Nam et al. \cite{nam2015clami}. The process is done on both the train and the test dataset. 
\be
\item For each $F_i$, go through all instances of $X_j$, a violation happens when $F_i$ at $X_j$ is higher than the $percentile(K_i, C)$ where $C = 50\%$ but $Y_j=1$ and vice-versa. 
\item Sum all the violations per feature across the whole dataset and sort it in ascending order.
\item Select the feature with the lowest violation score, if multiple of them have the same score then pick all of them. 
\ee 

\noindent \textbf{\underline{Instance Selection}}: 
\be
\item With the selected features, go through each instance $X_i$ and check if the respective $F_j$ values violated the proneness assumption then remove that instance $X_i$.  
\item If the dataset do not have instances with both classes at the end then pick the next minimum violation score to select metrics. 
\item This process is only done on the train dataset.
\ee

After selecting features with the minimum violation scores and removing the instances that violated the proneness tendency, a practitioner can train an RF model on the processed train data to identify the target classes from the processed test dataset. 

\section{Experimentation}

This section describes in details the experimental designs and the methodologies (i.e., standard semi-supervised learners from the machine learning community and the SOTA work from the additional software analytics).

\subsection{Experimental Design}

In order to make sure the method's effectiveness is not affected by the bias between deterministic and non-deterministic models or the bias of uncertainty, we
randomly shuffle training/test sets and incorporate stratified sampling with five bins (ensuring that the class distribution of the whole
data is replicated in each bin). The process is repeated for the training data but also includes an extra 2.5\% validating partition for each 97.5\% tuning partition. During the simulation, the tune partition will not review labels for the semi-supervised learning model. FRUGAL  does have access to the corresponding 2.5\% labelled validation partition while deciding on the best configurations. For each 20\% of the data (test set), the process learns a model on five stratified samples of the train data. This process is done for all four domains in this paper. \\

\subsection{Standard Semi-Supervised Learners}\label{standardssl}

\noindent \textit{\textbf{Self Training}} is simple and popular in recent SE research~\cite{yu2019improving, Yu19, tu_debtfree21, tu_frugal21, tu2020better}. It is based on Yarowsky et al.'s algorithm~\cite{self_training}. At first, a supervised classifier (here random forest) is trained on the  \textit{L\% randomly labeled set} and then incrementally unlabeled data points are predicted. For each iteration, the data points having prediction probability more than ``probability\_threshold'' (e.g., 0.7) are selected and added to the training set with the predicted labels. This process can be terminated when a certain criteria is fulfilled, e.g., max\_iteration is reached. Finally, as a result, we get a new training dataset which contains \textit{L\% randomly labeled set} and pseudo-labeled data points (by self-training). Such an advantage of using self-training is that any supervised classifier with good calibration can be used as the baseline model. \\

\noindent \textit{\textbf{Label Propagation}} is a semi-supervised graph inference algorithm as the ``community detection algorithm'' class (i.e., detect network structures or communities in a graph). Zhu et al~\cite{Zhu02learningfrom} developed the core algorithm. The algorithm starts with building a graph from the available labeled and unlabeled data. Each data point is a node in the graph and edges are the similarity weights. It uses a `kernel' function to project data into alternate dimensional spaces. The process will first assign unique labels to each node in the network. Then for each iteration, each node updates its label to the one that the maximum number of its neighbours belongs to. Generally, this process `propagates' the labels throughout the network and form communities~\cite{Zhu02learningfrom,LPADapeng}. Ties are broken using uniform logic:  
\begin{align*}
        C_{x}(t) = f(C_{x1}(t-1), C_{x2}(t-2), ....., C_{xk}(t-k))
    \end{align*}
At the end of traversing through the network for each iteration, the convergence condition is checked. If not met, the process continues to iterate.  \\

\noindent \textit{\textbf{Label Spreading}} was invented by Zhou et al.~\cite{label_spread}. It is also a graph inference algorithm like label propagation but has some differences. The method of changing the ground truth labels is called ``clamping''. Label propagation uses ``hard clamping'' as the original labels of the labeled points never changed. Meanwhile, label spreading is a ``soft clamping'' approach and more robust to noise as it adjusts the relative amount of information the points obtain from the neighbors. Hence, label spreading does not blindly believe the original labels (like label propagation) and makes modifications to the ground truth. The label spreading algorithm iterates on a modified version of the original graph and normalizes the edge weights by computing the normalized graph Laplacian matrix, i.e., symmetric normalized Laplacian matrix versus random walk normalized Laplacian matrix. \\

\noindent \textit{\textbf{Co-Training}:} is a very popular semi-supervised approach developed by Blum et al.~\cite{co_training}. Co-training has had great success in the text mining~\cite{Understand_cotraining,10.5555/3104322.3104466} and computer vision domain~\cite{BIANCO201615,qiao2018deep}. Co-training trains two classifiers based on two different views of labeled data, the feature set is divided into two mutually exclusive sets. Then both classifiers predict on the unlabeled data (e.g., \textit{clf1} and \textit{clf2}). The data points confidently predicted by \textit{clf1} are used for \textit{clf2} training, and vice versa. The unlabeled samples that are predicted with the most confident by \textit{clf1} and \textit{clf2} will be added to the training-set with the predicted label.  The success of co-training depends on a very specific assumption, ``original feature set can be divided into two mutually exclusive subsets which are conditionally independent given the class.''   \\

\subsubsection{State-of-the-art Methodologies}

\noindent\textit{\textbf{Support Vector Machine (SVM)}}: is a  classifier  defined by a separating hyperplane~\cite{suykens1999least}. Soft-margin linear SVMs are commonly used in text classification given the high dimensionality of the feature space. This was recommended by \citet{intrinsic_static} as the state of the art for our adoptable static code warning identification domain. A soft-margin linear SVM looks for the decision hyperplane that maximizes the margin between training data of two classes while minimizing the training error (hinge loss): \vspace{-10pt}
\begin{equation}
\label{eq:SVM}
\min {\lambda \lVert w\rVert ^{2}+\left[{\frac {1}{n}}\sum _{i=1}^{n}\max \left(0,1-y_{i}(w\cdot x_{i}-b)\right)\right]}
\end{equation}
where the class of $x$ is predicted as $sgn(w\cdot x-b)$.

\noindent\textit{\textbf{SWIFT}}: \citet{Shu2021HowTB} solved the dual optimization problems to optimize both learner and pre-processor options. SWIFT leveraged on the Deb et al.~\cite{deb2005evaluating}'s idea of $\epsilon$-dominance from 2005. \textit{$\epsilon$-dominance}  partitions the output space of an optimizer into $\epsilon$-sized grids. The principle of this idea is that if there exists some $\epsilon$ value below which it is useless or impossible to distinguish the results, then it superfluous to explore anything less than $\epsilon$. From a high level, SWIFT is essentially a tabu search; i.e., if some settings resulted in some performance within $\epsilon$ of any older result, then SWIFT marked that option as ``to be avoided''. SWIFT applied ``item ranking'' in seeking optimal learner and pre-processor, and further refines their option ranges. SWIFT returned the best setting seen during the following three stage process:

\bi
\item {\em Initialization}: all option items $i$ (learners/pre-processor) are assigned equal weightings.
\item The {\em item ranking} stage reweights items; e.g. terms like ``Random Forest'' or ``RobustScaler''.
\item The {\em numeric refinement} stage adjusts the tuning ranges of hyperparameter options of learners/pre-processor. \\ 
\ei

\noindent\textit{\textbf{FAIR\_SMOTE}}: Chakraborty et al.~\cite{fair_smote} proposed this method to achieve more balanced data, i.e., balancing both frequencies of not only class
labels but also sensitive attributes. The tradition SMOTE only focuses on the class labels. Prior work~\cite{chen19,fair_smote} in software fairness claimed that training data needs to be balanced in order to achieve fair prediction. Generally, training data is divided into four groups (Favorable \& Privileged, Favorable \& Unprivileged, Unfavorable \& Privileged, Unfavorable \& Unprivileged). Initially, these subgroups' sizes are not equal. After synthetic oversampling, the training data becomes balanced based on target class and protected attribute, i.e., above mentioned four groups become of equal sizes.

\section{Results}

\textbf{RQ1: Can we still reduce the cost of labeling for static warning analysis?} Due to data duplication and features leakage, static warning analysis's data from Tu et al.~\cite{tu_frugal21} is problematic as discussed in \S. Thanks to Kang et al.~\cite{kang_static}, the updated data is utilized here to revalidate FRUGAL's effectiveness against the SOTA for adoptable static warning identification. 

\citet{intrinsic_static} recommended off-the-shelf SVM as the SOTA solution for identifying adoptable static warnings. Table~\ref{tbl:swa} reports the comparison of FRUGAL against SVM across three metrics (i.e., recall, FAR, and AUC) on the updated dataset. We observe:

\bi
\item FRUGAL perform similarly to the SOTA at AUC, but outperforms in recall by 433\% relatively while reducing FAR by 30.7\% relatively.
\item In term of labeling efforts, FRUGAL costs 2.5\%, and the \citet{Shu2021HowTB}'s method costs 100\% because FRUGAL and the SOTA need 2.5\% and 100\% of the data labels to execute.  
\ei 

\begin{RQ}{\normalsize{In summary, our answer to RQ1 is:}} 
FRUGAL exceeds the SOTA SVM (EMSE'20 \cite{intrinsic_static} in identifying adoptable static warnings). FRUGAL requires only 2.5\% at median of the train data to be labeled  while using 97.5\% less information than the SOTA SVM. This validates and confirms FRUGAL's effectiveness for this task from the previous study. 
\end{RQ} \vspace{-10pt}

\textbf{RQ2: How well does FRUGAL distinguish security bug reports?}  

\begin{table*}[!t]
\footnotesize
\caption{Comparison of FRUGAL($L=2.5\%$) on three metrics as the SOTA SVM work (i.e., recall, FAR, and AUC) on median and number of wins across the aggregated dataset (across 8 projects). FRUGAL outperforms the SOTA SVM significantly in FAR and recall while performing similarly in AUC. }
\centering
\setlength\tabcolsep{5pt}
\begin{tabular}{l|c|c|c}
Treatment   & \multicolumn{1}{c|}{\cellcolor[HTML]{7DFFEA}\textbf{Recall}} &  \multicolumn{1}{c|}{\cellcolor[HTML]{7DFFEA}\textbf{FAR}} &  \multicolumn{1}{c}{\cellcolor[HTML]{7DFFEA}\textbf{AUC}}  \\ \hline

    FRUGAL\_2.5$\%$ &  \cellcolor[HTML]{FFCCC9}96\% & \cellcolor[HTML]{FFCCC9}9\% & \cellcolor[HTML]{FFCCC9}51\%\\
    SVM~\cite{intrinsic_static} & 18\% & 13\% & \cellcolor[HTML]{FFCCC9}52\% \\  \hline
    \multicolumn{1}{c|}{\textbf{\textit{M}}} & 19\% & 1\%  & 1\%  \\ 

\hline
\end{tabular}
\label{tbl:swa}
\end{table*}

\begin{table*}[!b]
\footnotesize
\caption{Comparison of FRUGAL($L \in \{2.5\%, 5\%, 10\%, 20\%\}$) on three metrics as the SOTA work (i.e., Recall, FAR, and IFA) on overal median and number of wins across 40 datasets (5 projects and 8 filters).  Except Recall, the lower the results the better the performance of the treatment. }
\centering
\scriptsize
\setlength\tabcolsep{3pt}
\begin{tabular}{l|c|c|c|c|c|c}
   & \multicolumn{2}{|c|}{\cellcolor[HTML]{7DFFEA}\textbf{Recall}} &  \multicolumn{2}{|c}{\cellcolor[HTML]{7DFFEA}\textbf{FAR}} &  \multicolumn{2}{|c}{\cellcolor[HTML]{7DFFEA}\textbf{IFA}}  \\ \hline
\multicolumn{1}{c|}{\cellcolor[HTML]{7DFFEA}\textbf{Treatment}}  & Wins & Median  & Wins  & Median & Wins  & Median \\ \hline
    FRUGAL\_2.5$\%$ & 10 & 31\% & 28 & 3\% & 23 & 4\\
    FRUGAL\_5$\%$  & 17 & 57\% & 29 & 3\% & 32 & 2\\
    FRUGAL\_10$\%$ & 26 & 85\% & 35 & 2\% & 27 & 3\\
    FRUGAL\_20$\%$ & 39 & 100\% & 38 & 0\% & 31 & 1\\  

\hline
\end{tabular}
\label{tab:sbr_FRUGAL}
\end{table*}

First, we start by investigating the amount of labeled data ($L\%$) right for FRUGAL in distinguishing security bug report  by vary $L\%$ values from 2.5\% up to 20\%. 

\begin{table*} 
\centering
\caption{{\bf RQ2} results. Except for Recall, the lower the results the better the performance of the treatment. For each row, the \colorbox[HTML]{FFCCC9}{highlighted} cells show best performing treatments (i.e., these cells that are statistically the same as the best median result -- as judged by our Cohen test). Across all rows, FRUGAL has the most number of best results.}
\setlength\tabcolsep{3pt}
\footnotesize
\begin{tabular}{l|l|c|c|c|c|c|c}
\multicolumn{2}{c|}{} &  \multicolumn{2}{|c|}{\cellcolor[HTML]{7DFFEA}\textbf{Recall}} &  \multicolumn{2}{|c}{\cellcolor[HTML]{7DFFEA}\textbf{FAR}} &  \multicolumn{2}{|c}{\cellcolor[HTML]{7DFFEA}\textbf{IFA}}  \\ \hline
\multicolumn{1}{c|}{\cellcolor[HTML]{7DFFEA}\textbf{Project}} & \multicolumn{1}{c|}{\cellcolor[HTML]{7DFFEA}\textbf{Filter}} & SWIFT & FRUGAL & SWIFT  & FRUGAL & SWIFT  & FRUGAL \\ \hline

 & train & 86 & \cellcolor[HTML]{FFCCC9}100 & 24 & \cellcolor[HTML]{FFCCC9}0 & \cellcolor[HTML]{FFCCC9}58 & 80 \\  
 & farsecsq & \cellcolor[HTML]{FFCCC9}72 & \cellcolor[HTML]{FFCCC9}66 & 14 & \cellcolor[HTML]{FFCCC9}0 & 36 & \cellcolor[HTML]{FFCCC9}0 \\  
 & farsectwo & 77 & \cellcolor[HTML]{FFCCC9}100 & 26 &  \cellcolor[HTML]{FFCCC9}0 & 87 & \cellcolor[HTML]{FFCCC9}0 \\  
 & farsec & \cellcolor[HTML]{FFCCC9}72  & \cellcolor[HTML]{FFCCC9}69 & 15 & \cellcolor[HTML]{FFCCC9}0 & \cellcolor[HTML]{FFCCC9}56 & 80 \\  
 & clni & 72 & \cellcolor[HTML]{FFCCC9}100 & 26  & \cellcolor[HTML]{FFCCC9}0 & 74 &  \cellcolor[HTML]{FFCCC9}64 \\  
 & clnifarsecsq & \cellcolor[HTML]{FFCCC9}86 & 52 & 14 & \cellcolor[HTML]{FFCCC9}0 & 37 & \cellcolor[HTML]{FFCCC9}0 \\ 
 & clnifarsectwo & 75 & \cellcolor[HTML]{FFCCC9}100 & 19 & \cellcolor[HTML]{FFCCC9}0 & \cellcolor[HTML]{FFCCC9}58 & 80    \\
\multirow{-8}{*}{Chromium} & clnifarsec & 75 & \cellcolor[HTML]{FFCCC9}100 & 20 &  \cellcolor[HTML]{FFCCC9}0 & \cellcolor[HTML]{FFCCC9}54 & \cellcolor[HTML]{FFCCC9}62    \\\cline{2-8} 
\multicolumn{2}{r|}{\textit{Median}} & 75 & 100 & 20 & 0 & 57 & 63    \\
 \multicolumn{6}{r}{~}\\\hline
 
 & train & \cellcolor[HTML]{FFCCC9}67 & 17 & 28 & \cellcolor[HTML]{FFCCC9}1 & \cellcolor[HTML]{FFCCC9}46 & \cellcolor[HTML]{FFCCC9}43  \\  
 & farsecsq & \cellcolor[HTML]{FFCCC9}83 & 17 & 67  & \cellcolor[HTML]{FFCCC9}6 & \cellcolor[HTML]{FFCCC9}39 & \cellcolor[HTML]{FFCCC9}41 \\  
 & farsectwo & \cellcolor[HTML]{FFCCC9}67 & 17 & 62 & \cellcolor[HTML]{FFCCC9}9& 31 &  \cellcolor[HTML]{FFCCC9}0  \\  
 & farsec & \cellcolor[HTML]{FFCCC9}33 & 17 & 23 & \cellcolor[HTML]{FFCCC9}8 & 22 & \cellcolor[HTML]{FFCCC9}0 \\  
 & clni & \cellcolor[HTML]{FFCCC9}50 & 17 & 14 & \cellcolor[HTML]{FFCCC9}4 & \cellcolor[HTML]{FFCCC9}27 & 45 \\  
 & clnifarsecsq & \cellcolor[HTML]{FFCCC9}83 & 17 & 58 & \cellcolor[HTML]{FFCCC9}0 & \cellcolor[HTML]{FFCCC9}6 & \cellcolor[HTML]{FFCCC9}0  \\  
 & clnifarsectwo & \cellcolor[HTML]{FFCCC9}67 & 17 & 26 & \cellcolor[HTML]{FFCCC9}0 & \cellcolor[HTML]{FFCCC9}8 & 45\\  
\multirow{-8}{*}{Wicket} & clnifarsec & \cellcolor[HTML]{FFCCC9}50 & 17 & 22 & \cellcolor[HTML]{FFCCC9}4 & \cellcolor[HTML]{FFCCC9}18 & 45 \\\cline{2-8} 
\multicolumn{2}{r|}{\textit{Median }} & 67 & 17 & 40 & 4 & 25 & 42\\
 \multicolumn{6}{r}{~}\\\hline
 
 & train & \cellcolor[HTML]{FFCCC9}86 & 14 & 18 & \cellcolor[HTML]{FFCCC9}3 & \cellcolor[HTML]{FFCCC9}1  & \cellcolor[HTML]{FFCCC9}0 \\  
 & farsecsq & \cellcolor[HTML]{FFCCC9}86 & 57 & 24 & \cellcolor[HTML]{FFCCC9}3 & \cellcolor[HTML]{FFCCC9}3 & \cellcolor[HTML]{FFCCC9}5\\  
 & farsectwo & \cellcolor[HTML]{FFCCC9}86 & 57 & 20 & \cellcolor[HTML]{FFCCC9}3 & \cellcolor[HTML]{FFCCC9}3  & \cellcolor[HTML]{FFCCC9}4 \\  
 & farsec & 86 & \cellcolor[HTML]{FFCCC9}100 & 20 & \cellcolor[HTML]{FFCCC9}3 & \cellcolor[HTML]{FFCCC9}17 & 85 \\  
 & clni & 86 & \cellcolor[HTML]{FFCCC9}100 & 18 & \cellcolor[HTML]{FFCCC9}3 & \cellcolor[HTML]{FFCCC9}13 & \cellcolor[HTML]{FFCCC9}15 \\  
 & clnifarsecsq & 71 & \cellcolor[HTML]{FFCCC9}100 & 29 & \cellcolor[HTML]{FFCCC9}3 & \cellcolor[HTML]{FFCCC9}8 & \cellcolor[HTML]{FFCCC9}2 \\ 
 & clnifarsectwo & 86 & \cellcolor[HTML]{FFCCC9}100 & 23 & \cellcolor[HTML]{FFCCC9}3 & \cellcolor[HTML]{FFCCC9}11 & \cellcolor[HTML]{FFCCC9}15\\
\multirow{-8}{*}{Ambari} & clnifarsec & \cellcolor[HTML]{FFCCC9}86 & 57 & 19 & \cellcolor[HTML]{FFCCC9}3 & \cellcolor[HTML]{FFCCC9}17 & \cellcolor[HTML]{FFCCC9}20 \\\cline{2-8}
\multicolumn{2}{r|}{\textit{Median }} & 86 & 79 & 20 & 3 & 10 & 10 \\

 \multicolumn{6}{r}{~}\\\hline
 & train & \cellcolor[HTML]{FFCCC9}56 & 17 & 16 & \cellcolor[HTML]{FFCCC9}4  & \cellcolor[HTML]{FFCCC9}15 & \cellcolor[HTML]{FFCCC9}12\\  
 & farsecsq & \cellcolor[HTML]{FFCCC9}67  & 17 & 28 & \cellcolor[HTML]{FFCCC9}13 & \cellcolor[HTML]{FFCCC9}14 & \cellcolor[HTML]{FFCCC9}12\\  
 & farsectwo & \cellcolor[HTML]{FFCCC9}61 & 17 & 45 & \cellcolor[HTML]{FFCCC9}17 & \cellcolor[HTML]{FFCCC9}25 & \cellcolor[HTML]{FFCCC9}17\\  
 & farsec & \cellcolor[HTML]{FFCCC9}56 & 17 & 36 & \cellcolor[HTML]{FFCCC9}4 & \cellcolor[HTML]{FFCCC9}8 & \cellcolor[HTML]{FFCCC9}8\\  
 & clni & \cellcolor[HTML]{FFCCC9}50 & 17 & 25 & \cellcolor[HTML]{FFCCC9}7 & 30 & \cellcolor[HTML]{FFCCC9}5\\  
 & clnifarsecsq & \cellcolor[HTML]{FFCCC9}61 & 17 & 27 & \cellcolor[HTML]{FFCCC9}5 & \cellcolor[HTML]{FFCCC9}10 & \cellcolor[HTML]{FFCCC9}2\\ 
 & clnifarsectwo & \cellcolor[HTML]{FFCCC9}61 & 17 & 39 & \cellcolor[HTML]{FFCCC9}7 & 12 & \cellcolor[HTML]{FFCCC9}2 \\
\multirow{-8}{*}{Camel} & clnifarsec & \cellcolor[HTML]{FFCCC9}56 & 17 & 37 & \cellcolor[HTML]{FFCCC9}9  & 22 & \cellcolor[HTML]{FFCCC9}4 \\\cline{2-8}
\multicolumn{2}{r|}{\textit{Median }} & 58 & 17 & 32 & 7 & 15 & 7 \\
 \multicolumn{6}{r}{~}\\\hline

 & train & \cellcolor[HTML]{FFCCC9}69 & 52 & 20 & \cellcolor[HTML]{FFCCC9}3 & \cellcolor[HTML]{FFCCC9}2 & \cellcolor[HTML]{FFCCC9}2  \\  
 & farsecsq & \cellcolor[HTML]{FFCCC9}67 & 45 & \cellcolor[HTML]{FFCCC9}0  & 48 & \cellcolor[HTML]{FFCCC9}4 & \cellcolor[HTML]{FFCCC9}2\\  
 & farsectwo & \cellcolor[HTML]{FFCCC9}79  & 31 & 40 & \cellcolor[HTML]{FFCCC9}7 & \cellcolor[HTML]{FFCCC9}3 & \cellcolor[HTML]{FFCCC9}2\\  
 & farsec & \cellcolor[HTML]{FFCCC9}64  & 31 & 14 & \cellcolor[HTML]{FFCCC9}0 & \cellcolor[HTML]{FFCCC9}4 & \cellcolor[HTML]{FFCCC9}2\\  
 & clni & \cellcolor[HTML]{FFCCC9}69  & 21 & 26 & \cellcolor[HTML]{FFCCC9}0 & \cellcolor[HTML]{FFCCC9}3 & \cellcolor[HTML]{FFCCC9}3\\  
 & clnifarsecsq & \cellcolor[HTML]{FFCCC9}67 & 33 & \cellcolor[HTML]{FFCCC9}42 & \cellcolor[HTML]{FFCCC9}47 & \cellcolor[HTML]{FFCCC9}1 & \cellcolor[HTML]{FFCCC9}2\\ 
 & clnifarsectwo & \cellcolor[HTML]{FFCCC9}62 & 29 & 52 & \cellcolor[HTML]{FFCCC9}22 & \cellcolor[HTML]{FFCCC9}4 & \cellcolor[HTML]{FFCCC9}2 \\
\multirow{-8}{*}{Derby} & clnifarsec & \cellcolor[HTML]{FFCCC9}67 & 52 & 20 & \cellcolor[HTML]{FFCCC9}0 & \cellcolor[HTML]{FFCCC9}2 & \cellcolor[HTML]{FFCCC9}2\\\cline{2-8} 
\multicolumn{2}{r|}{\textit{Median }} & 67 & 32 & 24 & 13 & 3 & 2\\
\multicolumn{6}{r}{~}\\\hline 
\rowcolor[HTML]{7DFFEA}\multicolumn{2}{r|}{\textit{Overall Median}} & 67 & 31 & 24 & 3 & 15 & 4\\\hline
\end{tabular}
\label{tbl:sbr}
\end{table*}

\begin{table*}[!t]
\footnotesize
\caption{Comparison of FRUGAL($L \in \{2.5\%, 5\%, 10\%, 20\%\}$) on three metrics as the SOTA work (i.e., recall, FAR, and IFA) on overal median and number of wins across 40 datasets (5 projects and 8 filters). The relative improvement gradually decreases on recall and disappears on IFA as $L$ increases.}
\centering
\setlength\tabcolsep{5pt}
\begin{tabular}{l|c|c|c|c|c|c}
   & \multicolumn{2}{c|}{\cellcolor[HTML]{7DFFEA}\textbf{Recall}} &  \multicolumn{2}{c|}{\cellcolor[HTML]{7DFFEA}\textbf{FAR}} &  \multicolumn{2}{c}{\cellcolor[HTML]{7DFFEA}\textbf{IFA}}  \\ \hline
\multicolumn{1}{c|}{\cellcolor[HTML]{7DFFEA}\textbf{Treatment}}  & Wins & Median  & Wins  & Median & Wins  & Median \\

\hline

    FRUGAL\_2.5$\%$ &  \cellcolor[HTML]{FFCCC9}11 & 32\% & \cellcolor[HTML]{FFCCC9}39 & 2\% & \cellcolor[HTML]{FFCCC9}33 & 4\\
    SWIFT~\cite{Shu2021HowTB} & 31 & 67\% & 3 & 24\% & 31 & 15\\  

\hline
\end{tabular}
\label{tab:sbr_summary}
\end{table*}

Our hypothesis is ``there are few key data regions where extra data would lead to indistinguishable results''. We test the different amounts of the train data's labels that are required for FRUGAL's performance to plateaus. Let $L$ be 2.5\%, 5\%, 10\%, or 20\%, Table~\ref{tab:sbr_FRUGAL} reports FRUGAL's performance on  security bug reports categorization (i.e., IFA, FAR, and recall). 
These metrics are derived from the SOTA's evaluation metrics~\cite{Shu2021HowTB}. The comparison is summarized as the overall median and the number of wins across 40 available datasets (i.e., 5 projects and 8 filters) for security bug reports categorization. A win describes a methodology that performs the best per dataset statistically.

However, for security bug report identification, the performance increases as $L$ increases. The relative improvements from increasing $L=10\%$ to 20\% are  17.6\%, 100\%, and 67\%  while the initial improvements (from $L=2.5\%$ to 10\%) are 174\%, 33\%, and 25\% across recall, FAR, and IFA correspondingly. A possible interpretation here is that $L=10\%$ are more suitable  
for  security bugs reports categorization. However, briefly observing FRUGAL's performance across all $L$, we notice that any $L$ would suffice to perform similarly or even better than the SOTA work from this task. Hence, $L=2.5\%$ would be optimal to minimize the labeling effort.

The changing effect were observed to be absent in previous study's issue close time prediction task and this study's software fairness, this is highly likely due to the more balanced nature of the data's class distribution. The median distribution of the minority class within software fairness is 30\%. However, the data in security bug reports are more imbalanced (with a median of 5\% of the target class respectively). This is consistent with the motivations for oversampling and undersampling techniques for imbalanced data~\cite{chawla2002smote, agrawal2018better}. \\

SWIFT is employed as the SOTA solution for distinguishing security bug reports. \citet{Shu2021HowTB} compared the proposed methods against three metrics (i.e., IFA, FAR, and recall) to finalize SWIFT as the best method for security bug reports categorization. Hence, Table~\ref{tbl:sbr} reports the comparison of FRUGAL against SWIFT on the same three metrics on each dataset while Table~\ref{tab:sbr_summary} summarizes Table~\ref{tbl:sbr} with overall number of wins and medians. We observe:

\bi
\item FRUGAL, on average, lost to relative SOTA's  recall by -52.2\% while reducing FAR and IFA by 91.3\% and 73.3\% relatively.
\item Among 40 datasets, FRUGAL statistically performs the best 11, 39, and 33 times for recall, FAR, and IFA respectively.
\item In term of labeling efforts, FRUGAL costs 2.5\%, and the \citet{Shu2021HowTB}'s method costs 100\% because FRUGAL and the SOTA need 2.5\% and 100\% of the data labels to execute.  
\ei

FRUGAL exceeds the SOTA SWIFT (EMSE'20 \cite{Shu2021HowTB}) in distinguishing security bug reports. FRUGAL requires only 2.5\% of the train data to be labeled when being compared against unsupervised learning while using 97.5\% less information than the SOTA tuned deep learning method. Hence, FRUGAL is also effective in distinguishing security bug reports.  \\

\begin{RQ}{\normalsize{In summary, our answer to RQ2 is:}} 
FRUGAL exceeds the SOTA SWIFT (EMSE'21 \cite{Shu2021HowTB} in distinguishing security bug reports). FRUGAL requires only 2.5\% at median of the train data to be labeled  while using 97.5\% less information than the SOTA SWIFT. This analysis gives insights to when FRUGAL does not perform well. 
\end{RQ} \vspace{-10pt}

\noindent \textit{\textbf{RQ3: Can FRUGAL better mitigate biases in SE ML models with
less data? }}

In this task, we first investigate the amount of labeled data ($L\%$) for FRUGAL across software fairness by vary $L\%$ values from 2.5\% up to 20\%.

\begin{table*}[!t]
\scriptsize
\caption{Comparison of FRUGAL($L \in \{2.5\%, 5\%, 10\%, 20\%\}$) on five metrics as the SOTA work (i.e., F1, AOD, EOD, DI, and SPD) on overall median and number of wins across 11 cases. Except F1, the lower the results the better the performance of the treatment.}
\centering
\setlength\tabcolsep{3pt}
\begin{tabular}{l|c|c|c|c|c|c|c|c|c|c}
   & \multicolumn{2}{c|}{\cellcolor[HTML]{7DFFEA}\textbf{F1}} &  \multicolumn{2}{|c}{\cellcolor[HTML]{7DFFEA}\textbf{AOD}} &  \multicolumn{2}{|c}{\cellcolor[HTML]{7DFFEA}\textbf{EOD}}  &  \multicolumn{2}{|c}{\cellcolor[HTML]{7DFFEA}\textbf{DI}} &  \multicolumn{2}{|c}{\cellcolor[HTML]{7DFFEA}\textbf{SPD}}\\ \hline
\multicolumn{1}{c|}{\cellcolor[HTML]{7DFFEA}\textbf{Treatment}}  & Wins  & Median   & Wins  & Median & Wins  & Median & Wins  & Median & Wins  & Median \\ \hline
    FRUGAL\_2.5$\%$ & 11 & 71\% & 9 & 4\% & 11 & 3\% & 5 & 6\%  & 7 & 2\%\\
    FRUGAL\_5$\%$  & 9 & 71\% & 9 & 3\% & 7 & 5\%  & 7 & 4\%  & 8 & 3\%\\
    FRUGAL\_10$\%$ & 10 & 71\% & 8 & 3\% & 7 & 3\%  & 10 & 4\%  & 8 & 3\%\\
    FRUGAL\_20$\%$ & 9 & 71\% & 10 & 3\% & 7 & 3\%  & 9 & 1\%  & 9 & 2\%\\  

\hline
\end{tabular}
\label{tab:fairness_FRUGAL}
\end{table*}

We conjecture that the same intuition of ``there are few key data regions where extra data would lead to indistinguishable results'' is also applicable here. Let $L$ be 2.5\%, 5\%, 2.5\%, or 20\%, Table~\ref{tab:fairness_FRUGAL} reports FRUGAL's performance on software fairness (i.e., F1, AOD, EOD, DI, and SPD). 
Again, these metrics are derived from the SOTA's evaluation metrics~\cite{fair_smote}. The comparison is summarized as the overall median and the number of wins across 11 available cases for software fairness. A win describes a methodology that performs statistically best per dataset.

\begin{table}[!t]
\scriptsize
\vspace{-5pt}
\caption{Comparison between  \textbf{FAIR\_SMOTE}\cite{fair_smote} and FRUGAL across five performance metrics (Recall, FAR, Precision, Accuracy, and F1) and four fairness metrics (AOD, EOD, SPD, and DI)    for software fairness. Note that in this table, the FRUGAL results were  obtained after labeling just  2.5\% of the data.
Except for Recall, Precision, Accuracy, and F1, the lower the results the better the performance of the treatment. Medians are calculated for easy comparisons. Here, the \colorbox[HTML]{FFCCC9}{highlighted} cells show best performing treatments (i.e., these cells that are statistically the same as the best median result -- as judged by our Cohen test).}
\vspace{-10pt}
\centering
\setlength\tabcolsep{3pt}
\renewcommand{\arraystretch}{1.1}

\begin{tabular}{c|l|c|c|c|c|c|c|c|c|c|c|c|c}
  \rowcolor[HTML]{7DFFEA}\textbf{Metrics}  & \textbf{Treatment} & \textbf{\rotatebox[origin=c]{90}{Adult (Sex)}} & \textbf{\rotatebox[origin=c]{90}{Adult (Race)}} & \textbf{\rotatebox[origin=c]{90}{Compas (Sex)}} &
  \textbf{\rotatebox[origin=c]{90}{Compas (Race)}} & \textbf{\rotatebox[origin=c]{90}{German (Sex)}} & \textbf{\rotatebox[origin=c]{90}{Default Credit (Sex)}} &
  \textbf{\rotatebox[origin=c]{90}{Heart-Health (Age)}} & \textbf{\rotatebox[origin=c]{90}{Bank Marketing (Age)}} & \textbf{\rotatebox[origin=c]{90}{Student (Sex)}} & \textbf{\rotatebox[origin=c]{90}{MEPS-15 (Race)}} & \textbf{\rotatebox[origin=c]{90}{MEPS-16 (Race)}} &
  \textbf{\rotatebox[origin=c]{90}{Median}}  \vspace{2pt}\\
\hline

Recall & FRUGAL	&	\cellcolor[HTML]{FFCCC9}99	&	\cellcolor[HTML]{FFCCC9}99	&	\cellcolor[HTML]{FFCCC9}100	&	\cellcolor[HTML]{FFCCC9}100	&	\cellcolor[HTML]{FFCCC9}100	&	\cellcolor[HTML]{FFCCC9}98	&	\cellcolor[HTML]{FFCCC9}99	&	\cellcolor[HTML]{FFCCC9}97	&	\cellcolor[HTML]{FFCCC9}100	&	\cellcolor[HTML]{FFCCC9}100	&	\cellcolor[HTML]{FFCCC9}100	&	100\\
$(M=6\%)$  &  FAIR\_SMOTE	&	71	&	70	&	62	&	62	&	62	&	58	&	66	&	76	&	91	&	68	&	66	&	66 \\ \hline								
FAR & FRUGAL	&	\cellcolor[HTML]{FFCCC9}2	&	\cellcolor[HTML]{FFCCC9}2	&	\cellcolor[HTML]{FFCCC9}0	&	\cellcolor[HTML]{FFCCC9}0	&	\cellcolor[HTML]{FFCCC9}0	&	\cellcolor[HTML]{FFCCC9}0	&	\cellcolor[HTML]{FFCCC9}11	&	\cellcolor[HTML]{FFCCC9}2	&	\cellcolor[HTML]{FFCCC9}1	&	\cellcolor[HTML]{FFCCC9}0	&	\cellcolor[HTML]{FFCCC9}0	&	0 \\
$(M=4\%)$  &  FAIR\_SMOTE	&	25	&	22	&	32	&	30	&	36	&	26	&	20	&	18	&	18	&	\cellcolor[HTML]{FFCCC9}4	&	\cellcolor[HTML]{FFCCC9}2	&	22 \\ \hline
													
Precision & FRUGAL	&	\cellcolor[HTML]{FFCCC9}68	&	\cellcolor[HTML]{FFCCC9}68	&	52	&	\cellcolor[HTML]{FFCCC9}52	&	31	&	\cellcolor[HTML]{FFCCC9}63	&	\cellcolor[HTML]{FFCCC9}80	&	\cellcolor[HTML]{FFCCC9}78	&	\cellcolor[HTML]{FFCCC9}100	&	\cellcolor[HTML]{FFCCC9}55	&	\cellcolor[HTML]{FFCCC9}46	&	63 \\
$(M=6\%)$  &  FAIR\_SMOTE	&	51	&	51	&	\cellcolor[HTML]{FFCCC9}56	&	\cellcolor[HTML]{FFCCC9}72	&	\cellcolor[HTML]{FFCCC9}71	&	39	&	69	&	\cellcolor[HTML]{FFCCC9}72	&	84	&	41	&	\cellcolor[HTML]{FFCCC9}41	&	56 \\ \hline
												
Accuracy & FRUGAL	&	\cellcolor[HTML]{FFCCC9}84	&	\cellcolor[HTML]{FFCCC9}84	&	\cellcolor[HTML]{FFCCC9}71	&	\cellcolor[HTML]{FFCCC9}71	&	\cellcolor[HTML]{FFCCC9}76	&	\cellcolor[HTML]{FFCCC9}80	&	\cellcolor[HTML]{FFCCC9}76	&	\cellcolor[HTML]{FFCCC9}73	&	81	&	\cellcolor[HTML]{FFCCC9}83	&	\cellcolor[HTML]{FFCCC9}83	&	80 \\
$(M=3\%)$  &  FAIR\_SMOTE	&	73	&	72	&	55	&	55	&	64	&	68	&	68	&	\cellcolor[HTML]{FFCCC9}72	&	\cellcolor[HTML]{FFCCC9}87	&	77	&	77	&	72 \\ \hline
												
F1 & FRUGAL	&	\cellcolor[HTML]{FFCCC9}71	&	\cellcolor[HTML]{FFCCC9}71	&	50	&	50	&	49	&	\cellcolor[HTML]{FFCCC9}80	&	\cellcolor[HTML]{FFCCC9}87	&	\cellcolor[HTML]{FFCCC9}79	&	\cellcolor[HTML]{FFCCC9}100	&	\cellcolor[HTML]{FFCCC9}72	&	\cellcolor[HTML]{FFCCC9}57	&	71 \\
$(M=5\%)$  &  FAIR\_SMOTE	&	62	&	62	&	\cellcolor[HTML]{FFCCC9}65	&	\cellcolor[HTML]{FFCCC9}66	&	\cellcolor[HTML]{FFCCC9}71	&	44	&	66	&	\cellcolor[HTML]{FFCCC9}74	&	86	&	53	&	51	&	65 \\ \hline
												
AOD & FRUGAL	&	\cellcolor[HTML]{FFCCC9}1	&	\cellcolor[HTML]{FFCCC9}2	&	6 	&	\cellcolor[HTML]{FFCCC9}4	&	\cellcolor[HTML]{FFCCC9}0	&	\cellcolor[HTML]{FFCCC9}2	&	27	&	\cellcolor[HTML]{FFCCC9}4	&	7	&	\cellcolor[HTML]{FFCCC9}3	&	\cellcolor[HTML]{FFCCC9}2	&	7 \\
$(M=2\%)$  &  FAIR\_SMOTE	&	\cellcolor[HTML]{FFCCC9}1	&	\cellcolor[HTML]{FFCCC9}4	&	\cellcolor[HTML]{FFCCC9}2	&	\cellcolor[HTML]{FFCCC9}1	&	5	&	\cellcolor[HTML]{FFCCC9}2	&	\cellcolor[HTML]{FFCCC9}8	&	\cellcolor[HTML]{FFCCC9}5	&	\cellcolor[HTML]{FFCCC9}4	&	\cellcolor[HTML]{FFCCC9}2	&	\cellcolor[HTML]{FFCCC9}1	&	2 \\ \hline
												
EOD & FRUGAL	&	\cellcolor[HTML]{FFCCC9}2	&	\cellcolor[HTML]{FFCCC9}2	&	\cellcolor[HTML]{FFCCC9}2	&	\cellcolor[HTML]{FFCCC9}4	&	\cellcolor[HTML]{FFCCC9}0	&	\cellcolor[HTML]{FFCCC9}4	&	30	&	\cellcolor[HTML]{FFCCC9}5	&	\cellcolor[HTML]{FFCCC9}4	&	\cellcolor[HTML]{FFCCC9}1	&	\cellcolor[HTML]{FFCCC9}2	&	2 \\
$(M=2\%)$  &  FAIR\_SMOTE	&	\cellcolor[HTML]{FFCCC9}2	&	\cellcolor[HTML]{FFCCC9}3	&	5	&	\cellcolor[HTML]{FFCCC9}5	&	13	&	\cellcolor[HTML]{FFCCC9}3	&	\cellcolor[HTML]{FFCCC9}7	&	\cellcolor[HTML]{FFCCC9}7	&	\cellcolor[HTML]{FFCCC9}4	&	\cellcolor[HTML]{FFCCC9}2	&	\cellcolor[HTML]{FFCCC9}3	&	4 \\ \hline
												
SPD & FRUGAL	&	\cellcolor[HTML]{FFCCC9}0	&	\cellcolor[HTML]{FFCCC9}2	&	\cellcolor[HTML]{FFCCC9}1	&	\cellcolor[HTML]{FFCCC9}8	&	\cellcolor[HTML]{FFCCC9}0	&	\cellcolor[HTML]{FFCCC9}2	&	27	&	\cellcolor[HTML]{FFCCC9}7	&	28	&	\cellcolor[HTML]{FFCCC9}1	&	\cellcolor[HTML]{FFCCC9}5	&	2 \\
$(M=3\%)$  &  FAIR\_SMOTE	&	\cellcolor[HTML]{FFCCC9}3	&	\cellcolor[HTML]{FFCCC9}5	&	8	&	\cellcolor[HTML]{FFCCC9}6	&	\cellcolor[HTML]{FFCCC9}5	&	\cellcolor[HTML]{FFCCC9}5	&	\cellcolor[HTML]{FFCCC9}8	&	\cellcolor[HTML]{FFCCC9}5	&	\cellcolor[HTML]{FFCCC9}4	&	5	&	\cellcolor[HTML]{FFCCC9}4	&	5 \\ \hline
												
DI & FRUGAL	&	\cellcolor[HTML]{FFCCC9}1	&	\cellcolor[HTML]{FFCCC9}11	&	\cellcolor[HTML]{FFCCC9}7	&	\cellcolor[HTML]{FFCCC9}11	&	\cellcolor[HTML]{FFCCC9}0	&	\cellcolor[HTML]{FFCCC9}0	&	6	&	10	&	19	&	\cellcolor[HTML]{FFCCC9}0	&	\cellcolor[HTML]{FFCCC9}0	&	6 \\
$(M=2\%)$  &  FAIR\_SMOTE	&	12	&	26	&	\cellcolor[HTML]{FFCCC9}9	&	\cellcolor[HTML]{FFCCC9}11	&	5	&	3	&	\cellcolor[HTML]{FFCCC9}2	&	\cellcolor[HTML]{FFCCC9}3	&	\cellcolor[HTML]{FFCCC9}8	&	15	&	17	&	9 \\
\hline
\end{tabular}
\label{tab:fairness}
\vspace{-15pt}
\end{table}

The observation from the software fairness is similar to the original FRUGAL study on static analysis and issue close time. Except for DI, FRUGAL's performance do not increase significantly as the $L$ increases. They even decrease on F1 and EOD (11 wins from $L=2.5\%$ on both metrics to 9 and 7 wins respectively). Hence, $L=2.5\%$ would be optimal for FRUGAL in this domain.

FAIR\_SMOTE is employed as the SOTA solution for mitigating biases in machine learning software. \citet{fair_smote} compared the proposed methods against five performance metrics (i.e., recall, precision, false-alarm rate, accuracy, and F1) and four fairness metrics (i.e., AOD, EOD, SPD, and DI) to finalize FAIR\_SMOTE as the best method for software fairness. Hence, Table~\ref{tab:fairness} reports the comparison of FRUGAL against FAIR\_SMOTE on the same nine metrics. We observe:

\bi
\item Across the five performance metrics, FRUGAL outperforms the SOTA FAIR\_SMOTE significantly, i.e., 8/11 times per metric. The relative median improvements are 9\%, 11\%, 12.5\%, 50\%, and 100\% for F1, accuracy, precision, recall, and FAR respectively.
\item Across the four fairness metrics, FRUGAL loses in AOD, draws in SPD, and wins in EOD and DI.
\item In term of labeling efforts, FRUGAL costs 2.5\%, and the \citet{Shu2021HowTB}'s method costs 100\% because FRUGAL and the SOTA need 2.5\% and 100\% of the data labels to execute.  
\ei

FRUGAL exceeds the SOTA FAIR\_SMOTE (FSE'21 \cite{Shu2021HowTB}) in software fairness. FRUGAL slightly outperforms FAIR\_SMOTE across the fairness metrics but dominates FAIR\_SMOTE across the performance metrics. Moreover, FRUGAL requires only 2.5\% of the train data to be labeled while using 97.5\% less information than the SOTA tuned deep learning method. Hence, FRUGAL is also effective in mitigating biases among ML software. The success in total of four areas let this study hypothesizes that other areas of SE may also benefit from FRUGAL.



\begin{RQ}{\vspace{-30pt}\normalsize{In summary, our answer to RQ3 is:}} \vspace{-2pt}
Even with non-SE data, FRUGAL still exceeds the SOTA FAIR\_SMOTE (FSE'21 \cite{fair_smote} in managing fairness in machine learning software). FRUGAL requires only 2.5\% at median of the train data to be labeled when being compared against unsupervised learning while using 97.5\% less information than both SOTA methods. Hence, FRUGAL is not only effective in static warning analysis and issue close time prediction, but also software fairness and security bug report. The success in these areas let this study hypothesizes that other areas of SE may also benefit from FRUGAL.
\end{RQ} \vspace{-16pt}

\noindent \textit{\textbf{RQ4: What labeling method would we recommend for SE data?}}

Another part of this generalization extension includes employing standard semi-supervised learners from the ML community in order to check if they suffice or even outperform our proposed method, FRUGAL. We conjecture that FRUGAL that was designed with domain knowledge of software analytics in mind (i.e., SE's data complexity) will outperform these standard methods. In order to validate our conjecture, we will conduct experiments to compare these standard SSL methods against FRUGAL across four previous tasks: adoptable static warning identification, issue close time prediction, security bug report categorization, and ML software fairness management.

\noindent \textit{\textbf{RQ4.a: How ML's standard semi-supervised learning methods perform in identifying adoptable static warnings?}}

For this first task, the four standard semi-supervised learners - self-training (ST), co-training (CT), label-spreading (LS), and label-propagation (LP) - are compared against FRUGAL across three metrics from prior work (i.e., FAR, recall, and AUC). Table~\ref{tab:chap6_swa_mlssl} reports that comparison. We observe:

\begin{table*}[!t]
\footnotesize
\caption{Comparison between self-training (ST), co-training (CT), label spreading (LS), label propagation (LP), and FRUGAL in terms of FAR, Recall, and AUC for identifying adoptable static warning.
In this table, all results were  found after labeling just  2.5\% of the data.
Except for FAR, the higher the results the better the performance of the treatment. Here, the \colorbox[HTML]{FFCCC9}{highlighted} cells show best performing treatments.}
\centering
\setlength\tabcolsep{5pt}
\begin{tabular}{l|c|c|c}
Treatment   & \multicolumn{1}{c|}{\cellcolor[HTML]{7DFFEA}\textbf{Recall}} &  \multicolumn{1}{c|}{\cellcolor[HTML]{7DFFEA}\textbf{FAR}} &  \multicolumn{1}{c}{\cellcolor[HTML]{7DFFEA}\textbf{AUC}}  \\ \hline
    ST &  13.4\% & 10.5\% & 46.4\%\\
    LS &  0\% & \cellcolor[HTML]{FFCCC9}0\% & 47.8\%\\
    LP &  0\% & \cellcolor[HTML]{FFCCC9}0\% & 47.8\%\\ 
    CT &  26.4\% & 14.3\% & 44.8\%\\ 
    FRUGAL\_2.5$\%$ &  \cellcolor[HTML]{FFCCC9}95.5\% & 9.2\% & \cellcolor[HTML]{FFCCC9}49.8\%\\ \hline
    \multicolumn{1}{c|}{\textbf{\textit{M}}} & 13.9\% & 2.3\%  & 0.7\%  \\

\hline
\end{tabular}
\label{tab:chap6_swa_mlssl}
\end{table*}

\bi

\item Among standard SSL methods, LP and LS outperform ST and CT in AUC and FAR while underperforming in recall. This indicates that graph-based SSL methods are more effective than the other twos for this task.

\item Similarly, FRUGAL performs the best as FRUGAL only loses in FAR but wins in recall and AUC. When comparing against the highest performing standard SSL method, FRUGAL's absolute improvements are -9.2\%, 2\%, and 69.1\%  on FAR, AUC, and recall respectively. 

\item Regarding the labeling efforts, all methods are learned on 2.5\% data labels so the comparison is fair and there is no cost saving here.  

\ei

FRUGAL outshines standard ML's SSL methods for adoptable static warning identification. This indicates the rudimentary effectiveness of the SE data complexity intuition within FRUGAL that standard ML's SSL methods do not have access to. \\

\begin{table}[!t]
\footnotesize
\vspace{-5pt}
\caption{Comparison between self-training (ST), co-training (CT), label spreading (LS), label propagation (LP), and FRUGAL in terms of FAR, Recall, and AUC for predicting issues close time. Note that in this table, all results were  obtained after labeling just  2.5\% of the data.
Except for FAR, the higher the results the better the performance of the treatment. Medians and IQRs (delta between 75th
and 25th percentile, lower the better) are calculated for easy comparisons. Here, the \colorbox[HTML]{FFCCC9}{highlighted} cells show best performing treatments.}
\vspace{-5pt}
\centering
\setlength\tabcolsep{5pt}
\renewcommand{\arraystretch}{1.1}

\begin{tabular}{c|l|c|c|c|c|c}
  \textbf{Metrics}  & \textbf{Treatment} & \textbf{\rotatebox[origin=c]{90}{Chromium}} & \textbf{\rotatebox[origin=c]{90}{Firefox}} & \textbf{\rotatebox[origin=c]{90}{Eclipse}} & \textbf{\rotatebox[origin=c]{90}{Median}} & \textbf{\rotatebox[origin=c]{90}{IQR}} \vspace{2pt}\\
\hline
    	&	ST	&	\cellcolor[HTML]{FFCCC9}65	&	69.7	&	\cellcolor[HTML]{FFCCC9}66.1	&	66.1	&	2.35 \\
    	&	LS	&	56.5	&	59.6	&	57.2	&	57.2	&	1.55 \\
    Accuracy 	&	LP	&	56.5	&	60.2	&	57.6	&	57.6	&	1.85 \\
    ($M = 1.9\%$) 	&	CT	&	\cellcolor[HTML]{FFCCC9}63.5	&	68.6	&	64.4	&	64.4	&	2.5 \\
	&	FRUGAL	&	\cellcolor[HTML]{FFCCC9}65.3	&	\cellcolor[HTML]{FFCCC9}74.2	&	\cellcolor[HTML]{FFCCC9}68.2	&	68.2	&	4.5 \\

\hline
& CT	&		68.5	&		76.1	&		67.7 & 68.5 & 4.2 \\
& ST	&		70.8	&		78.3	&		70.7 & 70.8 & 3.8 \\
F1 & LS	&		60.7	&		67.3	&		58.6 & 60.7 &	4.4 \\
(M=2.6\%) & LP	&		60.6	&		67.8	&		58.3 & 60.6 &	4.8 \\
& FRUGAL	&		\cellcolor[HTML]{FFCCC9}75	&		\cellcolor[HTML]{FFCCC9}82.3	&		\cellcolor[HTML]{FFCCC9}74.8 & 75	& 3.8 \\

\hline

		&	ST	&	51	&	54.2	&	50.3	&	51	&	1.95  \\
    	&	LS	&	54.2	&	54.8	&	51.2	&	54.2	&	1.8 \\
    AUC	&	LP	&	57	&	57	&	54.8	&	57	&	1.1 \\
    ($M = 3.5\%$)  	&	CT	&	52.1	&	58.6	&	47.3	&	52.1	&	5.65 \\
	& 	FRUGAL	&	\cellcolor[HTML]{FFCCC9}72.1	&	\cellcolor[HTML]{FFCCC9}80.2	&	\cellcolor[HTML]{FFCCC9}75.8	&	75.8	&	4.1 \\
\hline
\end{tabular}
\label{tab:chap6_ict_mlssl}
\vspace{-15pt}
\end{table}

\noindent \textit{\textbf{RQ4.b: How ML's standard semi-supervised learning methods perform in predicting issue close times?}}

For this second task, the four standard semi-supervised learners - self-training (ST), co-training (CT), label-spreading (LS), and label-propagation (LP) - are compared against FRUGAL across three metrics from prior work (i.e., accuracy, FAR, recall, and AUC). Table~\ref{tab:chap6_ict_mlssl} reports that comparison. We observe:

\bi

\item Among standard SSL methods, ST performs the best by winning in accuracy, recall, and FAR while losing in AUC. The graph-based SSL methods, LS and LP, only win in FAR and recall. 

\item FRUGAL performs the best across all four metrics and all datasets. When comparing against the highest performing standard SSL method, FRUGAL's relative improvements are 3.2\%, 27.1\%, 48.6\%, and 96\%  on accuracy, recall, AUC, and FAR respectively. 

\item Regarding the labeling efforts, all methods are learned on 2.5\% data labels so the comparison is fair and there is no cost saving here.  

\ei 

FRUGAL also outshines standard ML's SSL methods for issue close time prediction. This confirms the  effectiveness of the SE data complexity intuition within FRUGAL that standard ML's SSL methods do not have access to. 

Interestingly, graph-based methods (LS and LP) that were observed more effective than the other twos for adoptable static warnings identification task are not effective for issue close time prediction. This is possibly due to the data balance nature of this task that are absent in adoptable static warnings identification  and security bug report  categorization tasks. Moreover, ST performs similarly to the previous SOTA SIMPLE by winning on accuracy and recall while losing on FAR and AUC with only 10\% of the data's labels. \\

\noindent \textit{\textbf{RQ4.c: How ML's standard semi-supervised learning methods perform in distinguishing security bug reports?}}

For the third task, the four standard semi-supervised learners - self-training (ST), co-training (CT), label-spreading (LS), and label-propagation (LP) - are compared against FRUGAL across three metrics from prior work (i.e., FAR, recall, and IFA). Table~\ref{tab:sbr_mlssl} reports that comparison. We observe:

\bi

\item Among standard SSL methods, ST and CT barely outperform LS and LP by winning in IFA and recall while underperforming in FAR. Because of the lack of clear evidences, the difference between SSL methods are inconclusive.

\item FRUGAL performs the best as FRUGAL barely loses in FAR but wins in recall and IFA. When comparing against the highest performing standard SSL method, FRUGAL's absolute improvements are -2.1\%, 3\%, and 84.6\% on FAR, IFA, and recall respectively. 

\item Regarding the labeling efforts, all methods are learned on 10\% data labels so the comparison is fair and there is no cost saving here.  

\ei 

\begin{table*}[!t]
\footnotesize
\caption{Comparison between self-training (ST), co-training (CT), label spreading (LS), label propagation (LP), and FRUGAL in terms of FAR, Recall, and AUC for distinguishing security bug reports. Note that in this table, all results were  obtained after labeling just  2.5\% of the data.
Except for FAR, the higher the results the better the performance of the treatment. Overall Medians are calculated for easy comparisons. Here, the \colorbox[HTML]{FFCCC9}{highlighted} cells show best performing treatments.}
\centering
\setlength\tabcolsep{5pt}
\begin{tabular}{c|l|c|c|c|c|c|c|c}
   \cellcolor[HTML]{7DFFEA}\textbf{Metrics} & \cellcolor[HTML]{7DFFEA}\textbf{Treatment}  & \cellcolor[HTML]{7DFFEA}\textbf{\rotatebox[origin=c]{90}{AMBARI}} & \cellcolor[HTML]{7DFFEA}\textbf{\rotatebox[origin=c]{90}{CAMEL}} & \cellcolor[HTML]{7DFFEA}\textbf{\rotatebox[origin=c]{90}{CHROMIUM}} & \cellcolor[HTML]{7DFFEA}\textbf{\rotatebox[origin=c]{90}{DERBY}} & \cellcolor[HTML]{7DFFEA}\textbf{\rotatebox[origin=c]{90}{WICKET}} & \cellcolor[HTML]{7DFFEA}\begin{tabular}[c]{@{}c@{}}\textbf{Overall}\\ \textbf{Wins}\end{tabular} & \cellcolor[HTML]{7DFFEA}\begin{tabular}[c]{@{}c@{}}\textbf{Overall}\\ \textbf{Median}\end{tabular} \vspace{2pt}\\ 

\hline
    	&	ST    &    0    &    0    &    0     &    19.1    &    0    &    1    &    0 \\
    	&	LS    &    0    &    0    &    0    &    0    &    0    &    0    &    0 \\
    Recall	&	LP    &    0    &    0    &    0    &    0    &    0    &    0    &    0 \\
    ($M = 11.5\%$)	&	CT	&	0    &    0    &    0.5    &    20.3    &    0    &    1    &    0   \\  
	& 	FRUGAL	& 	\cellcolor[HTML]{FFCCC9}100    &    \cellcolor[HTML]{FFCCC9}17    &    \cellcolor[HTML]{FFCCC9}79    &    \cellcolor[HTML]{FFCCC9}17    &    \cellcolor[HTML]{FFCCC9}32    &    40    &    31 \\

\hline
		&	ST	&	\cellcolor[HTML]{FFCCC9}0    &   \cellcolor[HTML]{FFCCC9} 0    &    \cellcolor[HTML]{FFCCC9}0    &   \cellcolor[HTML]{FFCCC9} 0.9    &   \cellcolor[HTML]{FFCCC9} 0    &    38    &    0 \\
    	&	LS	&	\cellcolor[HTML]{FFCCC9}0    &    \cellcolor[HTML]{FFCCC9}0    &    \cellcolor[HTML]{FFCCC9}0    &    \cellcolor[HTML]{FFCCC9}0    &    \cellcolor[HTML]{FFCCC9}0    &    40    &    0 \\
    FAR	&	LP	&	\cellcolor[HTML]{FFCCC9}0    &    \cellcolor[HTML]{FFCCC9}0    &    \cellcolor[HTML]{FFCCC9}0    &    \cellcolor[HTML]{FFCCC9}0    &    \cellcolor[HTML]{FFCCC9}0    &    40    &    0 \\
    ($M = 4.1\%$)     &   CT    &    \cellcolor[HTML]{FFCCC9}0    &    \cellcolor[HTML]{FFCCC9}0    &    \cellcolor[HTML]{FFCCC9}0    &    \cellcolor[HTML]{FFCCC9}2.2    &    \cellcolor[HTML]{FFCCC9}0    &    38    &    0 \\
	  & 	FRUGAL	&  		\cellcolor[HTML]{FFCCC9}0    &    \cellcolor[HTML]{FFCCC9}4    &    \cellcolor[HTML]{FFCCC9}3    &    7   &    13    &    32    &    3 \\

\hline
		&	ST    &    -1    &    -1    &    -1    &    -0.5    &    -1    &    1    &    -1 \\
    	&	LS    &    -1    &    -1    &    -1    &    -1    &    -1    &    0    &    -1  \\
    IFA  & 	 LP    &    -1    &    -1    &    -1    &    -1    &    -1    &    0    &    -1 \\
    ($M = 2.7\%$) & 	CT    &    -1    &    -1    &    -1    &    \cellcolor[HTML]{FFCCC9}0.5    &    -1    &    4    &    -1 \\
	& 	FRUGAL    &    \cellcolor[HTML]{FFCCC9}63    &    \cellcolor[HTML]{FFCCC9}42    &    \cellcolor[HTML]{FFCCC9}10    &    \cellcolor[HTML]{FFCCC9}7    &    \cellcolor[HTML]{FFCCC9}2    &    38    &    4 \\

\hline
\end{tabular}
\label{tab:sbr_mlssl}
\end{table*}

The underwhelming performance of standard SSL methods in distinguishing security bug reports can be traced back to two main reasons. Firstly, the target class proportion within the data are significantly low, only 5.1\% on average whereas 15\% in adoptable static warnings identification and 50\% in issue close time prediction. SE literature has documented that software analytics tools often struggle to excel with imbalanced SE datasets. Secondly, they only use 2.5\% of the dataset's labels which is only 10 data instances (the median count of dataset's bug reports number is 400) so there is less than 1 target instance or security bug reports (5.1\% of 10) among them. It is difficult for standard SSL methods without the SE domain knowledge to learn the insights from a tiny dataset with little target instances. Hence, they end up predicting everything as non-security bug reports which result in very low FAR and recall along with invaluable IFA (i.e., -1).

FRUGAL again outshines standard ML's SSL methods for security bugs report categorization. This indicates the  effectiveness of the SE data complexity intuition within FRUGAL that standard ML's SSL methods do not have access to. 

\noindent \textit{\textbf{RQ4.d: How ML's standard semi-supervised learning methods perform in managing fairness in machine learning software?}}

For the fourth task, the four standard semi-supervised learners - self-training (ST), co-training (CT), label-spreading (LS), and label-propagation (LP) - are compared against FRUGAL across five metrics from prior work. This includes one performance metric (i.e., F1) and four fairness metrics (i.e., AOD, EOD, SPD, DI). Table~\ref{tab:sbr_mlssl} reports that comparison. We observe:

\bi

\item Among standard SSL methods, ST outperforms the rest by a small margin. When being compared to the second best of LP, ST wins in AOD and SPD, draws in F1 and DI, and loses in EOD. LS and LP perform similarly (win 2, lose 2, draw 1) but LP's total number of wins is higher than LS by one. 

\item FRUGAL performs the best as FRUGAL wins across the five metrics. When comparing against the highest performing standard SSL method, it is notable that FRUGAL, at median, improves F1 by 51.1\% and reduces AOD and EOD by 42.9\% and 73\% relatively. 

\item Regarding the labeling efforts, all methods are learned on 2.5\% data labels so the comparison is fair and there is no cost saving here.  

\ei 

\begin{table*}[!t]
\scriptsize
\caption{Comparison of FRUGAL($L \in \{2.5\%, 5\%, 10\%, 20\%\}$) on five metrics as the SOTA work (i.e., F1, AOD, EOD, DI, and SPD) on overall median and number of wins across 11 cases. Except F1, the lower the results the better the performance of the treatment. Here, the \colorbox[HTML]{FFCCC9}{highlighted} cells show best performing treatments.}
\centering
\setlength\tabcolsep{3pt}
\begin{tabular}{l|c|c|c|c|c|c|c|c|c|c}
   & \multicolumn{2}{c|}{\cellcolor[HTML]{7DFFEA}\textbf{F1}} &  \multicolumn{2}{|c}{\cellcolor[HTML]{7DFFEA}\textbf{AOD}} &  \multicolumn{2}{|c}{\cellcolor[HTML]{7DFFEA}\textbf{EOD}}  &  \multicolumn{2}{|c}{\cellcolor[HTML]{7DFFEA}\textbf{DI}} &  \multicolumn{2}{|c}{\cellcolor[HTML]{7DFFEA}\textbf{SPD}}\\ \hline
\multicolumn{1}{c|}{\cellcolor[HTML]{7DFFEA}\textbf{Treatment}}  & Wins  & Median   & Wins  & Median & Wins  & Median & Wins  & Median & Wins  & Median \\ \hline
    FRUGAL & \cellcolor[HTML]{FFCCC9}9 & 71\% & \cellcolor[HTML]{FFCCC9}8 & 4\% & \cellcolor[HTML]{FFCCC9}8 & 3\% & \cellcolor[HTML]{FFCCC9}8 & 6\%  & \cellcolor[HTML]{FFCCC9}8 & 2\%\\
    ST & 5 & 47\% & 6 & 7\% & 5 & 11.1\% & 7 & 5\% & 7 & 0\%\\
    LS & 4 & 57\% & 6 & 7.4\% & 5 & 9.4\% & 5 & 6\% & 5 & 19\%\\
    LP & 5 & 57\% & 5 & 4.2\% & 6 & 3.6\% & 7 & 5\% & 5 & 21\%\\
    CT & 3 & 45\% & 4 & 5.2\% & 5 & 7.1\% & 6 & 5\% & 6 & 3\%\\

\hline
\end{tabular}
\label{tab:fairness_ssl}
\end{table*}

Although, the performance of standard SSL methods cannot outperforms FRUGAL but they are a lot closer than the security bug reports case. This is due to the software fairness data is less imbalanced than the security bug reports (30\% versus 5\% respectively). FRUGAL again outshines standard ML's SSL methods for mitigating biases in ML software. This indicates the  effectiveness of the SE data complexity intuition within FRUGAL that standard ML's SSL methods do not have access to. 

From this RQ, we conclude that: 

\begin{RQ}{\normalsize{Result:}} 
FRUGAL outperforms all standard semi-supervised learning method from the machine learning community across four software analytics tasks. Moreover, this ineffectiveness is hypothetically due to the (1) highly imbalanced data nature  that is not often observed in standard ML tasks; and (2) these standard SSL methods does not tap into that SE domain knowledge that FRUGAL has access to. 
\end{RQ}

\section{Threats of Validity}\label{tion:threats}

There are several validity threats~\cite{feldt2010validity} to the design of this finalized solution (i.e., FRUGAL). Any conclusion made from this work must be considered with the following issues in mind:

\textbf{Conclusion validity} focuses on the significance of the treatment. To enhance   conclusion validity, we run experiments on 52 different  projects across stratified sampling (25 runs)  and find that our proposed method always performed better than the state-of-the-art approaches. More importantly, we apply a similar statistical testing of Cohen'd as the SOTA works of \citet{tu_frugal21}'s, to obtain fair comparison.  In addition, we have taken into generalization issues of single evaluation metrics (e.g., recall and precision) into consideration and instead evaluate our methods on metrics that aggregate multiple metrics like AUC while being effort-aware via cost.  As future work, we plan to test the proposed methods with additional analyses that are endorsed within SE literature (e.g., P-opt20~\cite{tu2020better}) or general ML literature (e.g., MCC~\cite{mcc_metrics}).

According to Tu et al.~\cite{tu_frugal21}, the simple effectiveness of both binary split of the output space (FRUGAL's centrality) and 2.5\% labeled train data requirement is highly due to how the approach is a strong heuristic that leverages on the intrinsic dimensionality. \citet{levina2004maximum} argued that many datasets embedded in high-dimensional spaces can be compressed without significant information loss (similar to the PCA method \cite{MACKIEWICZ1993303}).  To compute Levina's intrinsic dimensionality, a 2$-$d plot is created where the x-axis shows $r$; i.e., the radius of two configurations while the y-axis shows $C(r)$ as the number of configurations after spreading out some distance $r$ away from any of $n$ data instances:  
  
\begin{equation}
y = C(r) = {\frac{2}{n(n-1)}\sum\limits_{i=1}^n \sum\limits_{j=i+1}^n I\left[\lVert x_i, x_j \rVert < r \right]}
\end{equation}

The maximum slope of  $\ln C(r)$ vs. $\ln r$ is then reported as the intrinsic dimensionality, $D$.
Note that
$I[\cdot]$ is the indicator function (i.e., $I[x] = 1$ if $x$ is true, otherwise it is 0);
$x_i$ is the $i$th sample in the dataset. Applying this calculation to the 50+ datasets of the domain (reports in Table \ref{tab:intrinsic_dim}), we found the intrinsic or latent dimensionality ($D$) of our data is very low. The median $D$ are 2.06, 1.44, and 1.43 for static warnings, security bug reports, and software fairness respectively. FRUGAL's effectiveness essentially roots in how  the performance score generated from SE data can be divided into a few regions (low dimensional). FRUGAL's central function of binary splitting compresses the data dimensions (features) via aggregated percentile $C$ and survey the whole space by varying $C$ ($\{5\%$ to $95\%$ increments by $5\%\}$). Menzies et al.~\cite{menzies07} and Hindle et al.~\cite{hindle2012naturalness} also reported on how several SE data are low dimensional and the benefits from building effective tools from such data. This  work extends those findings: the labeling efforts to commission to tools building  can be reduced greatly even on non-SE data as long as they share the low dimensionality characteristic of SE data~\cite{hindle2012naturalness, agrawal2019dodge, menzies07, intrinsic_static, simple_ict}.  

\newcolumntype{Y}{>{\centering\arraybackslash}X}

\begin{table}[!t]
\small
\caption{Summary of intrinsic dimensions ($D$) of this study's 50 datasets from \citet{levina2004maximum}.}
\vspace{-5pt}
\centering
\setlength\tabcolsep{2pt}
\renewcommand{\arraystretch}{1.05}

\scriptsize
\begin{tabularx}{0.7\linewidth}{P{.09\linewidth}P{.15\linewidth}|P{.07\linewidth}|P{.07\linewidth}|P{.07\linewidth}|P{.07\linewidth}|P{.07\linewidth}}
\midrule
  &  & \multicolumn{5}{c}{\textbf{Security Bug Reports}} \\ 
 \cmidrule(lr){3-7} 
   \textbf{} & \textbf{Static Code Warnings} &  \textbf{\rotatebox[origin=c]{90}{Ambari}} & \textbf{\rotatebox[origin=c]{90}{Camel}} & \textbf{\rotatebox[origin=c]{90}{Chromium}}  & \textbf{\rotatebox[origin=c]{90}{Derby}}  & \textbf{\rotatebox[origin=c]{90}{Wicket}}  \vspace{3pt}\\

\midrule

\multicolumn{1}{c|}{\textbf{Median $D$}} & 2.06 & 1.62 & 1.2 & 1.52 & 1.43 & 0.98 \\

\bottomrule
\end{tabularx}

\begin{tabularx}{0.85\linewidth}{P{.09\linewidth}P{.07\linewidth}|P{.07\linewidth}|P{.07\linewidth}|P{.07\linewidth}|P{.07\linewidth}|P{.07\linewidth}|P{.07\linewidth}|P{.07\linewidth}|P{.07\linewidth}}
\midrule
  &   \multicolumn{9}{c}{\textbf{Software Fairness}} \\ 
 \cmidrule(lr){2-10} 
   \textbf{} &  \textbf{\rotatebox[origin=c]{90}{Adult}}&  \textbf{\rotatebox[origin=c]{90}{Compas}}&  \textbf{\rotatebox[origin=c]{90}{German}}&  \textbf{\rotatebox[origin=c]{90}{Default}} & \textbf{\rotatebox[origin=c]{90}{Heart Health}} & \textbf{\rotatebox[origin=c]{90}{Bank}}  & \textbf{\rotatebox[origin=c]{90}{Student}}  & \textbf{\rotatebox[origin=c]{90}{MEPS15}} & \textbf{\rotatebox[origin=c]{90}{MEPS16}}  \vspace{3pt}\\

\midrule

\multicolumn{1}{c|}{\textbf{$D$}} & 0.21 & 0.25 & 1.54 & 3.46 & 2.74 & 1.44 & 0.55 & 1.44 & 1.63 \\

\bottomrule
\end{tabularx}
\label{tab:intrinsic_dim}
\end{table}

\textbf{Construct validity }focuses on the relation between the theory behind the experiment and the observation. 
To enhance construct validity, we benchmarked our solution with the SOTA's solutions across several domains to ensure that our proposed solution's effectiveness is not due to random sampling of the data. However, we only show that with our default parameters settings of random forest learner. The performance can get even better by tuning the parameters, employing different learners (e.g., deep learners), and introducing a variety of data preprocessors (e.g., synthetic minority over-sampling or SMOTE that is known to help with imbalanced datasets \cite{chawla2002smote,agrawal2018better} like our static code warnings domains). We aim to explore these in our future work.

\textbf{Internal validity} focuses on how sure we can be that the treatment caused the outcome. To enhance   internal validity, we heavily constrained our experiments to the same dataset, with the same settings, except for the treatments being compared.

\textbf{External validity} concerns how widely our conclusions can be applied. In order to test the generalizability of our approach, we always kept a project as the holdout test set and never used any information from it in training. Moreover, we have validated our proposed method on four important software analytics domains: adoptable static code warnings identification, issues close time prediction, security bug report detection, and ML software fairness management. 

\textbf{Sampling Bias}, like any data mining work, this work is threatened by sampling bias; i.e. what holds for the data we studied here may
not hold for other kinds of data. For instance, for security bug report detection, FRUGAL outperformed the SOTA with only 2.5\% but the performance actually plateaus after $L=10\%$. However, FRUGAL's performance plateaus beyond $2.5\%$ data's labels for adoptable static warning identification, issue close time prediction, and software fairness. That said, and to repeat the message of this study, when different methods work for different data, researchers must take the time to carefully check the ground truth in the new data. Let that check be overwhelmingly slow and expensive, we recommend the use of semi-supervised learning with our proposed tool here, i.e.,  FRUGAL.

\section{Conclusion and Future Work}

Previous work introduced a general semi-supervised learning solution (i.e., FRUGAL) to reduce the labeling effort while keeping the performance similar or better in building software analytics tools. However, several problems associated with that work include: (1) only evaluated on two tasks while half of the previous work's conclusions were refuted due to data quality issues; (2) did not employ the method on non-SE data; and (3) did not benchmark against standard or ``weak'' SSL methods from the ML community. Therefore, in  this  paper,  we  have  used  FRUGAL  to  applying  this  ``strong''  heuristic (with SE domain knowledge)  to tackle those problems by: 

\be
\item For static code warnings, we revalidated and confirmed Tu et al.'s FRUGAL's effectiveness on the updated data. 
\item We extended FRUGAL's effectiveness from SE data to non-SE data, i.e., software fairness.
\item Our work identified two extra domains that also share data labeling effort and quality concerns, i.e., software fairness and security bug reports. These include more than 50 datasets, ten times more than the previous study and FRUGAL still outperformed both tasks while using 97.5\% less data or reducing the cost of labeling data by a factor of 40 times (100\%/2.5\%). 
\item Our and previous work conclusions were further generalized and strengthened by contrasting against the standard SSL methods from the machine learning community. Moreover, we also asserted that FRUGAL does best while benchmarking against standard SSL methods from the machine learning community as its strong heuristic capability to leverage on the SE domain knowledge. Essentially, specific SE knowledge is more useful (for SE applications) than domain-independent notions (e.g., similar things share similar properties in graph-inference algorithms).
\item This work demonstrates the benefits of Semi-Supervised Learning for Software Analytics. This includes software fairness, security bug reports, static warning analysis, and issue close time. This suggests that many more software analytics tasks could benefit from unsupervised learning and semi-supervised learning. As mentioned above, those benefits include the ability to commission new models with less human efforts and costs. By restricting human involvement in the process, we also reduced erroneous labels that can cascade to the whole research community since human are still error-prone (Yu et al.~\cite{jitterbug} found 98\% of the false-positive labels within \citet{maldonado2015detecting} were actually true-positive labels).
\item Overall, our study restated the benefit in exploring low dimensional SE data~\cite{hindle2012naturalness, agrawal2019dodge, menzies07, intrinsic_static, simple_ict, tu_frugal21} and extended their findings that the labeling efforts can be reduced greatly for non-SE data as long as they share the SE data's low dimensionality characteristics. 
\ee

That said, FRUGAL still suffers from the validity threats discussed in \S\ref{tion:threats}. To further reduce those threats and to move forward with this research, we propose the following future work:
\bi
\item Test whether replacing the Random Forest model in FRUGAL  with a deep learning model will further improve its performance.
\item Apply non-trivial hyper-parameter tuning (e.g., DODGE~\cite{agrawal2019dodge} or FLASH~\cite{nair2017flash}) on various data preprocessors and machine learners with FRUGAL to test whether tuning can further improve the performance.
\item
Extend the work to other software engineering domains (e.g., technical debts~\cite{maldonado2015detecting}, software configurations~\cite{Estublier}, etc) and compare it with other state-of-the-art methods which continue to appear.
\ei

\section*{Acknowledgements}
This work was partially funded by an NSF CISE Grant \#1931425.

\section*{Conflict of Interest}

The authors have no competing interests to declare that are relevant to the content of this article. 

\bibliographystyle{plainnat}
\bibliography{main.bbl}
 


\clearpage
\appendix

\end{document}